\documentclass[aip,jcp,notitlepage,10pt,twocolumn,superscriptaddress]{revtex4-1}
\usepackage[utf8]{inputenc}
\usepackage{geometry}
\usepackage{xcolor}

\usepackage{natbib}
\usepackage{graphicx}
\usepackage{amsmath}
\usepackage{graphicx}
\usepackage{wasysym}
\usepackage{stmaryrd}
\usepackage{caption}
\usepackage{subfigure}
\usepackage{float}
\begin{document}
\title{Characterization and Efficient Monte Carlo Sampling of Disordered Microphases}
\author{Mingyuan Zheng}
\affiliation{Department of Chemistry, Duke University, Durham, North Carolina 27708, United States}
\author{Patrick Charbonneau}
\affiliation{Department of Chemistry, Duke University, Durham, North Carolina 27708, United States}
\affiliation{Department of Physics, Duke University, Durham, North Carolina 27708, United States}
\date{\today}
\begin{abstract}
The disordered microphases that develop in the high-temperature phase of systems with competing short-range attractive and long-range repulsive (SALR) interactions result in a rich array of distinct morphologies, such as cluster, void cluster and percolated (gel-like) fluids. These different structural regimes exhibit complex relaxation dynamics with significant relaxation heterogeneity and slowdown. The overall relationship between structure and configurational sampling schemes, however, remains largely uncharted. In this article, the disordered microphases of a schematic SALR model are thoroughly characterized, and structural relaxation functions adapted to each regime are devised. The sampling efficiency of various advanced Monte Carlo (MC) sampling schemes--Virtual-Move (VMMC), Aggregation-Volume-Bias (AVBMC) and Event-Chain (ECMC)--is then assessed. A combination of VMMC and AVBMC is found to be computationally most efficient for cluster fluids and ECMC to become relatively more efficient as density increases. These results offer a complete description of the equilibrium disordered phase of a simple microphase former as well as dynamical benchmarks for other sampling schemes. 
\end{abstract}

\maketitle

\section{Introduction}
\label{sec:introduction}
Microphases are thermodynamically stable mesoscale structures found in diverse hard and soft materials, such as block copolymers \cite{riess2003micellization,bates2000block} cell nuclei\cite{MCMO19}, colloidal suspensions \cite{liu2019colloidal,seul1995domain}, surfactant solutions \cite{gelbart2012micelles,gelbart2012micelles} and magnetic alloys \cite{portmann2003inverse}. Their rich self-assembly behavior has notably found applications in drug delivery \cite{kataoka2012block,rosler2012advanced}, nanoscale patterning \cite{li2006block,krishnamoorthy2006nanoscale}, lithography \cite{hawker2005block,tang2008evolution} and filtration membranes.\cite{YQMMC15}  
A common microscopic origin for microphases is competing  short-range attractive and long-ranged repulsive (SALR) interactions, irrespective of the underlying material constituents \cite{stradner2004equilibrium,campbell2005dynamical,ciach2013origin,santos2017thermodynamic,CZ21}. Short-range attraction promotes condensation, while long-range repulsion frustrates the formation of extended domains. Turning on frustration thus first suppresses the gas-liquid critical point and eventually substitutes it (beyond the Lifshitz point) by a weakly first-order order-disorder transition (ODT) \cite{bates1990block,seul1995domain,matsen1996unifying,brazovskiui1996phase,bates2000block,zhuang2016recent}.

Microphases have aroused considerable material interests over the last decades \cite{leibler1980theory,zhuang2016equilibriumPRL,geissler2004nature,de2006columnar,verso2006star,ghezzi1997formation,sear1999spontaneous,archer2007phase,lindquist2016assembly,zhuang2017communication,toledano2009colloidal,sciortino2005one,mani2014equilibrium,godfrin2014generalized,jadrich2015origin}, yet some of their properties remain poorly understood. The facile formation of richly ordered microphases in block copolymers and its agreement with early theory \cite{leibler1980theory,bates1990block,matsen1996unifying} were foundational to the field, but such periodic structures, often fail to assemble. In colloidal suspensions, in particular, disordered microphases, such as cluster and percolated fluids are then typically observed~\cite{stradner2004equilibrium,campbell2005dynamical}, in defiance of theoretical expectations~\cite{ciach2013origin,zhuang2016equilibriumPRL,geissler2004nature,de2006columnar,verso2006star,ghezzi1997formation,sear1999spontaneous}. The potential of colloidal suspensions to form void aggregates (cavities or pores)~\cite{archer2007phase,lindquist2016assembly} akin to inverse structures of block copolymers or amphiphiles
\cite{matsen1996unifying,gelbart2012micelles} is also of putative interest for filtration and catalysis. Our limited understanding of disordered microphases in general, and of void-based structures in particular, however, hinders further development.

To make matters worse, numerical simulations of the disordered microphases are computationally challenging. Not only do they require relatively large system sizes for mesoscale structures to form and for spurious correlations to vanish, but sampling and equilibration are also fairly slow in this regime. Consider, for instance, cluster formation which, akin to micelle formation, occurs at low temperatures and beyond a concentration range known as critical clustering density (ccd) (by analogy to the critical micelle concentration (cmc)). In experiments, two main dynamical pathways are generally acknowledged~\citep{lang1975chemical,aniansson1974kinetics}: a fast particle exchange between clusters and environments, and a slow cluster formation and disintegration. Both are activated processes and the latter can be considerably slowed by high free energy barriers to cluster formation~\citep{nyrkova2005theory}. In this context, numerical simulations that rely on standard (local) sampling algorithms are inefficient in two respects. First, the energetic barrier to single particle displacements can be dynamically prohibitive. Second, both intra-cluster relaxation and whole-cluster diffusion rely on even slower collective rearrangements. These hurdles hinder parameter space exploration, and may even limit our confidence in conclusions drawn from certain numerical simulations.  In order to efficiently sample the long-range displacement of both clusters and particles as well as that of inter-cluster particle exchanges, advanced sampling techniques thus ought to be considered. Several generic candidate schemes exist--VMMC (Virtual-Move Monte Carlo) \cite{whitelam2007avoiding,whitelam2009role,ruuvzivcka2014collective}, AVBMC (Aggregation-Volume-Bias Monte Carlo) \cite{chen2000novel} and ECMC (Event-Chain Monte Carlo) \cite{michel2014generalized}--but their relative (and combined) efficacy and structural adequacy remains largely untested for microphase formers.

In this article, we conduct such a study by both thoroughly describing the various structural regimes of disordered microphases of a schematic SALR model, and by characterizing the dynamics of various advanced sampling schemes.
In what follows, we first describe the model studied  (Sec.~\ref{sec:model}), and the Monte Carlo simulation methods employed (Sec.~\ref{sec:MCmethods}), before introducing the relevant dynamical observables (Sec.~\ref{sec:observables}). Results of these observables for the simulation methods in various regimes are then discussed in  Sec.~\ref{sec:discussion}. A brief conclusion follows in Sec.~\ref{sec:conclusion}.

\section{Models and Characterization}
\label{sec:model}
\begin{figure}
\centering
\includegraphics[width=8.5cm,trim={0.8cm 1cm 0 1cm},clip]{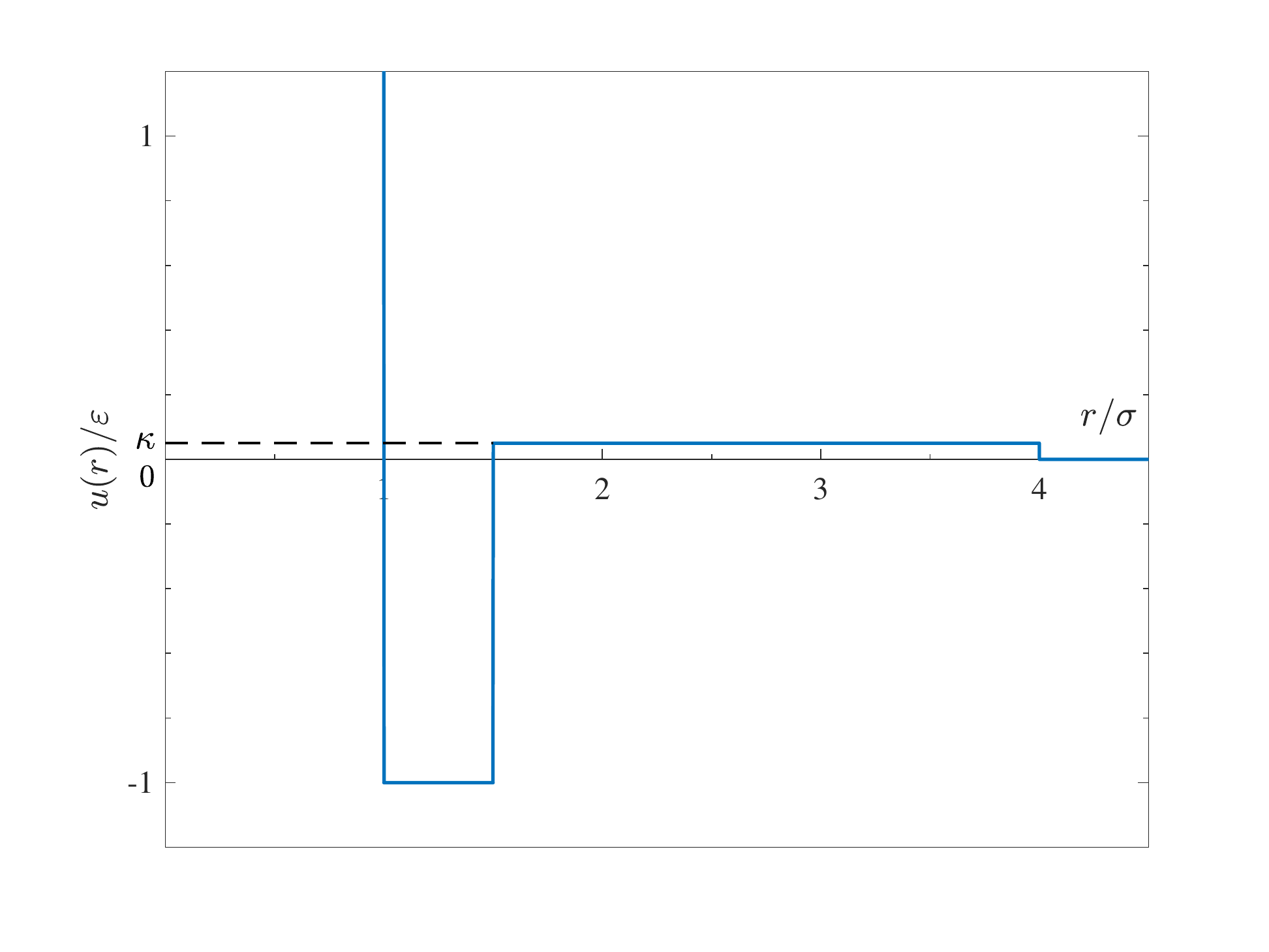}
\caption{Radial profile of the SALR pair interaction considered in this work. The hard core repulsion for $r < \sigma$, is followed by a square well attraction over $\sigma < r < \lambda_0\sigma$ and then by a constant repulsive step, extending over $\lambda_0\sigma < r < \lambda_1\sigma$.} 
\label{fig:salr}
\end{figure}

The pair interaction potential for the schematic SALR model considered in this work is given by (see Fig.~\ref{fig:salr})
\begin{equation}
\label{eq:hamiltonian}
     u(r) = \begin{cases}
              \infty & r < \sigma\\
             -\varepsilon & \sigma < r < \lambda_0\sigma\\
             \kappa\varepsilon & \lambda_0\sigma < r < \lambda_1\sigma\\
             0 & r>\lambda_1\sigma
             \end{cases},
\end{equation}
for which a hard-core of diameter $\sigma$, which sets the unit of length, is followed by a square-well attraction of strength $\varepsilon$, which sets the unit of energy. Attraction is felt up to $\lambda_0\sigma$, after which a repulsion of strength $\kappa\varepsilon$ is felt up to $\lambda_1\sigma$. This particular continuous space model is chosen so as to capture the essential material behavior of microphase formers, while also being amenable to various computational optimizations. Although not specifically designed to recapitulate any particular experimental system, it is thought to be reasonably representative of generic simple microphase formers. (In this context, the computational convenience of Eq.~\eqref{eq:hamiltonian} for examining the structures and dynamics of diverse regimes outweighs the considerations of legacy SALR models, including our own.)

In this work, we study systems of $N=250$-5000 particles, depending on the structural regime considered, in cubic boxes of volume $V$ under periodic boundary conditions. We also set $\lambda_0 = 1.5$, $\lambda_1 = 4.0$, $\kappa = 0.025$. The results and analysis are robust beyond this specific choice as long as: (i) the attraction range is taken to be broad enough to have a stable condensation transition in the absence of frustration, and thus keep sampling relatively facile; (ii) the repulsion range is taken to be broad enough to frustrate a few layers of neighbors yet short enough to not overly increase computational costs; (iii) the combination of frustration range and strength is sufficient to bring the system beyond the Lifshitz point.

%figure: potential

%figure: h(rho)
\begin{figure}[thbp]
\centering
\subfigure{
\begin{minipage}[t]{0.48\textwidth}
\centering
\includegraphics[width=8.5cm,trim={0.8cm 0.1cm 0 0.8cm},clip]{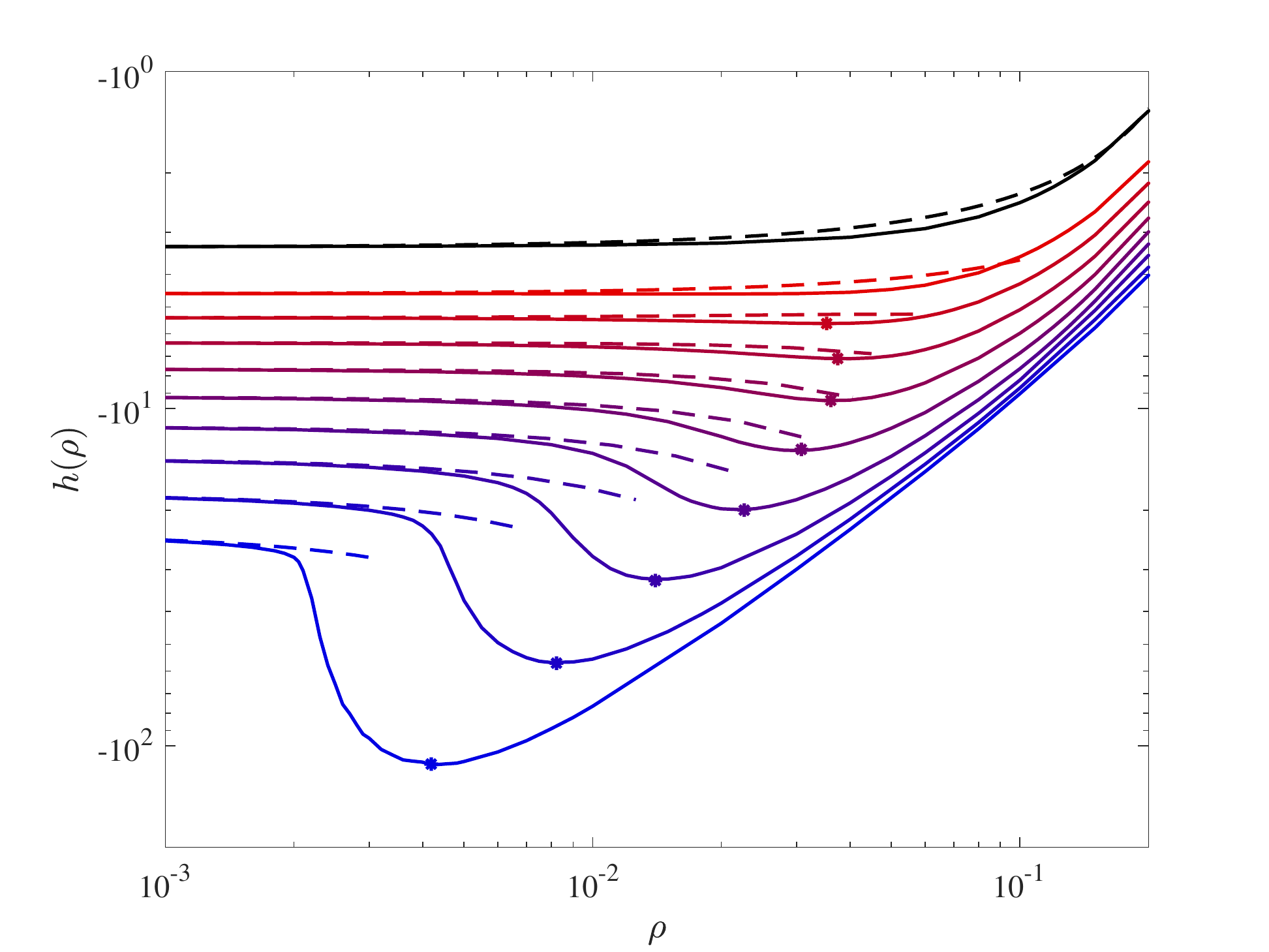}
\end{minipage}
}
\caption{Density evolution of the low-density reduced equation of state, Eq.~\eqref{eq:h_rho}, for $T = 0.5$-$0.9$ in increments of $0.05$ as well as for $T = 1.0$, from bottom to top (solid lines). The critical clustering density (ccd, stars) can be extracted from the first (local) minimum of $h(\rho)$, which vanishes between $0.85$ and $0.9$. As expected, the virial series (dashed lines) is consistent with simulation results (for $N=500$ using ECMC, see Sec.~\ref{sec:MCmethods}) at low densities. In particular, they have the same intercept ($B_2(T)$), initial slope ($B_3(T)$), and curvature ($B_4(T)$). The strongly cooperative nature of clustering, however, leads to marked deviations from the low-order virial series in the vicinity of the ccd.}
\label{fig:hrho}
\end{figure}

The simplicity of the chosen model also makes its first few virial coefficients -- $B_n(T)$ for temperature $T$ -- analytically accessible (Appendix~\ref{sec:appendix}). Their consideration is particularly instructive, because systems with SALR interactions are known to display significant deviations from ideality even at relatively small densities \cite{santos2017thermodynamic}. A proposed way to assess the onset of clustering indeed examines the deviations from the ideal gas pressure $P$ equation of state as \cite{zhuang2016equilibriumPRL, hu2018clustering}
\begin{equation}\label{eq:h_rho}
    h(\rho; T) = \frac{\beta P - \rho}{\rho^2} 
               = B_2(T) + B_3(T)\rho + B_4(T)\rho^2 + O(\rho^3),
\end{equation}
for number density $\rho=N/V$ and temperature $T$ (and its inverse $\beta=1/T$ with Boltzmann constant $k_B=1$)
and identifies the ccd as a marked depression in $h(\rho;T)$.  (A peak in the heat capacity offers another common estimate of the ccd \cite{frantz1995magic,imperio2006microphase,pekalski2019self,schwanzer2016two}, as further discussed in Sec.~\ref{sec:discussion}.) Comparing the virial series with $B_n(T)$ up to $n=4$  with simulation results reveals a near quantitative agreement at low densities, which cross-validates both approaches (Fig.~\ref{fig:hrho}). Once clustering emerges, however, very pronounced a deviation from the virial series arises. The cooperative nature of clustering is such that even at $\rho\sim 10^{-3}$, tens of virial coefficients (roughly, the typical cluster size) would likely be needed to recapitulate the simulation results. In other words, although clustering is not associated with any free energy singularity, it nevertheless corresponds to a very strong perturbation around the ideal gas limit.

\section{Advanced Monte Carlo Methods}
\label{sec:MCmethods}
This strong deviation from ideality, even under dilute conditions, further motivates the consideration of importance sampling methods in simulations.  In this section, we review various schemes that have previously been used, albeit on an \emph{ad hoc} basis, to sample the disordered microphases, namely Virtual-Move Monte Carlo (VMMC) \citep{whitelam2007avoiding,whitelam2009role,ruuvzivcka2014collective}, and Aggregation-Volume-Bias Monte Carlo (AVBMC) \citep{chen2000novel}. We also consider Event-Chain Monte Carlo (ECMC) \citep{michel2014generalized}, which has not been tested in this context, but might improve the sampling of disordered microphases. In all cases, we expect the sampling efficiency of these algorithms to be considerably better than that of standard single-particle (local) Metropolis Monte Carlo (MMC). In order to quantify this enhancement, we define an MC unit of time as attempting $N$ single-particle displacements, irrespective of the details of the sampling scheme. Note that collective moves may displace a significant fraction of the system, hence (artificially) decorrelating particle positions. To account for this artefact, a center of mass shift correction is maintained to properly compute the correlations described in Sec.~\ref{sec:observables}.

% figure: methods
\begin{figure}
\centering
\includegraphics[width=8cm,trim={0cm 0cm 0 0cm},clip]{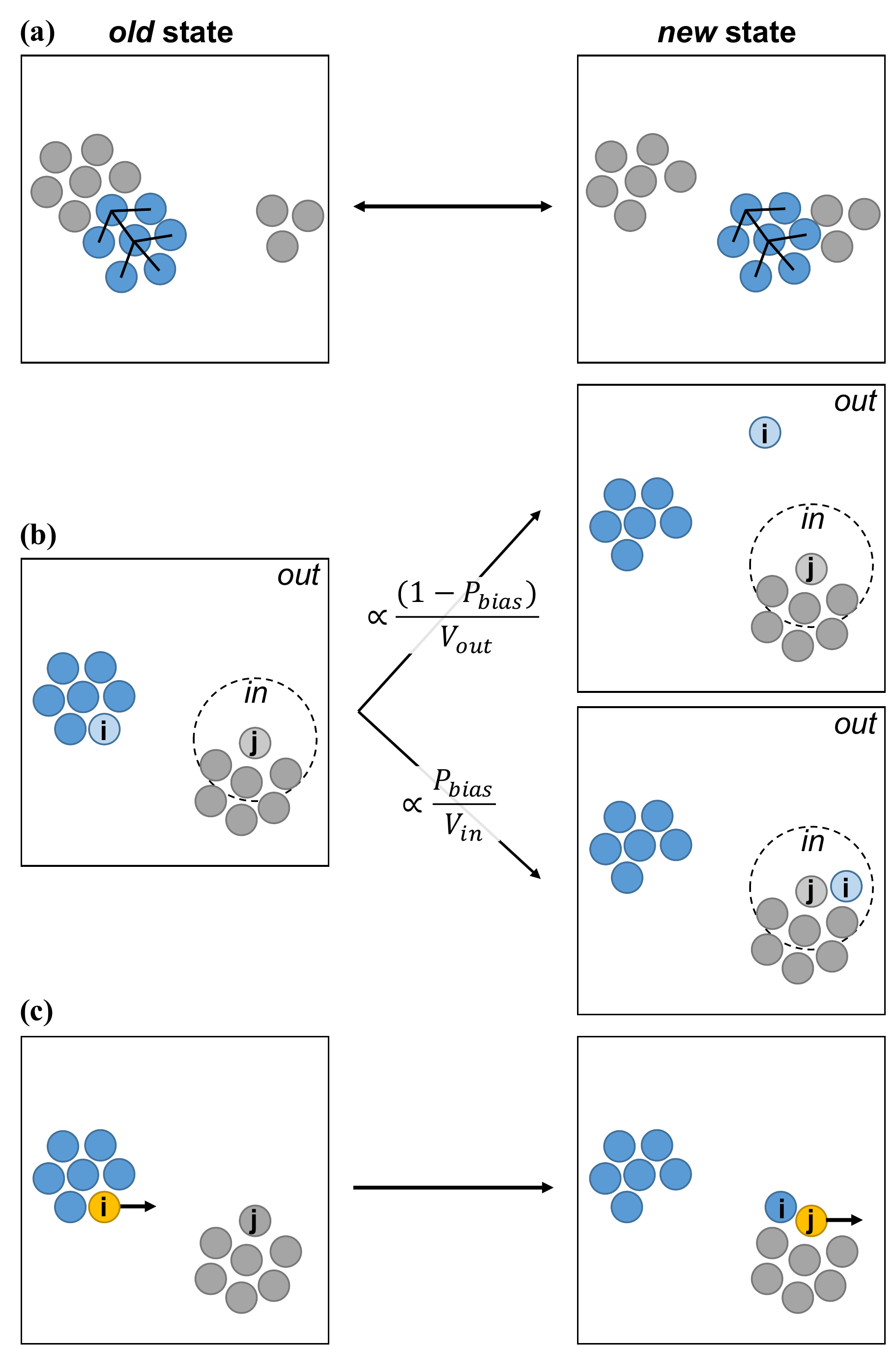}
\caption{Schematics of the various advanced Monte Carlo sampling algorithms considered in this work. (a) VMMC coherently displaces a pseudocluster (blue) obtained from an iterative linking scheme. Solid lines denote internal links. (b) AVBMC swaps a target particle $i$ to the bonded (\textit{in}) or unbonded (\textit{out}) region of another selected particle $j$ with a preassigned biasing probability. Only the case with $i$ initially \textit{out} of the neighboring of $j$ is shown here. (c) ECMC proceeds through \emph{lifting}, with a moving particle (orange) transferring motion to its collision partner.} 
\label{fig:methods}
\end{figure}

\subsection{Virtual-Move Monte Carlo}
Virtual-Move Monte Carlo is a type of cluster-move Monte Carlo \cite{whitelam2007avoiding} that simultaneously displaces a series of neighboring particles as a pseudocluster (see Fig.~\ref{fig:methods}(a)). The method is specifically designed to realize the collective displacement of groups of strongly attracting particles, while preserving microscopic reversibility. By applying collective (rather than single-particle) moves, VMMC accelerates both intra-cluster rearrangements and whole-cluster diffusion. 

More specifically, the approach randomly selects a particle to displace, and the collective motion of its neighbors is determined by a preset probability associated with the bonding strength,
\begin{equation}\label{eq: plink}
    p_{ij}(o\rightarrow n) = \left\{ \begin{array}{ll}
              p_\mathrm{link}, & \mbox{\ \ $r_{ij}(o) < \lambda_0\sigma$};\\
              0, & \mbox{\ \ $r_{ij}(o) \geq \lambda_0\sigma$}.\end{array} \right. .
\end{equation}
The constant $p_\mathrm{link}\in [0, 1)$ is optimized for a given pair potential. For the SALR model in Eq.~\eqref{eq:hamiltonian}, only particles within the range of attraction are considered as potentially linked. The pseudocluster is then grown by recursively testing the linking probability of neighbors of other cluster members. The acceptance ratio for the collective motion that follows from the detailed balance is then
\begin{align}
    &\frac{\mbox{acc}(o \rightarrow n)}{\mbox{acc}(n \rightarrow o)} \nonumber\\
    &= \min{\left\lbrace 1, e^{-\beta\Delta E} \cdot \prod\limits_{\langle ij\rangle_{u}}\frac{1 - p_{ij}(n \rightarrow o)}{1 - p_{ij}(o \rightarrow n)}\right\rbrace}
\end{align}
where $\Delta E$ is the energy difference between configurations $o$ and $n$, and $\langle ij\rangle_u$ denotes unlinked particle pairs, in which one particle is in the pseudocluster and the other is not.

For systems with large clusters--including elongated or even percolated structures--the size distribution of pseudoclusters is a particularly crucial consideration. Because the seed of the pseudocluster is randomly selected, larger clusters are more likely to be considered, and rejecting trial moves for these clusters is computationally wasteful. Pseudocluster displacements may then also span the entire system, and thus not significantly relax the overall system structure. To minimize this possibility, we follow the proposal of Ref.~[\onlinecite{whitelam2007avoiding}] of defining a cutoff of pseudocluster size $n_c$ with probability $\propto 1/n_c$ for each move, and thus each cluster can be selected with equal probability. If the size of the pseudocluster exceeds $n_c$, then the probability of forming other links (Eq.~\eqref{eq: plink}) is zero~\cite{ruuvzivcka2014collective}. A reduced rejection rate ensues.

Note that the rotational counterpart of this scheme was also considered, but was found not to significantly improve sampling, and is therefore not further discussed here.

\subsection{Aggregation-Volume-Bias Monte Carlo}
In systems with strong attraction (or, equivalently, at low temperatures), the local MMC moves inefficiently sample phase space because many particles are then similarly locally trapped. Specific configurations may then have a large Boltzmann weight because of their low energy, but their overall relative contribution to phase space remains small. In order to surmount this specific sampling challenge, the biased non-local Aggregation-Volume-Bias Monte Carlo (AVBMC) was proposed \cite{chen2000novel}. This scheme improves cluster sampling efficiency by moving one particle from one aggregate to another, regardless of the intermediate energy barrier (see Fig.~\ref{fig:methods}(b)). As long as most particles are at the surface of clusters, surface-to-surface displacements within or between clusters accelerate sampling.

More specifically, for AVBMC a randomly selected particle $i$ is either inserted into the bonded region of another random particle $j$ with probability $\propto p_\mathrm{bias}/V_\mathrm{in}$, or into the unbonded region with probability $\propto (1-p_\mathrm{bias})/V_\mathrm{out}$, where $p_\mathrm{bias}$ is the preset biasing probability, and $V_\mathrm{in}$ or $V_\mathrm{out}$ is the volume of inner (bonded) or outer (unbonded) region of particle $j$. Considering the relative positions of these two particles before and after displacements, the acceptance rules of the four resulting cases are,
\begin{equation}
\begin{split}
    \mbox{acc}(o_\mathrm{out}\rightarrow n_\mathrm{out}) &= \mbox{acc}(o_\mathrm{in}\rightarrow n_\mathrm{in}) \\
    &=\min{\left[1, e^{-\beta\Delta E}\right]}
\end{split}
\end{equation}
\begin{equation}
    \mbox{acc}(o_\mathrm{out}\rightarrow n_\mathrm{in}) =
    \min{\left[1, \frac{(1-p_\mathrm{bias})V_\mathrm{in}}{p_\mathrm{bias}V_\mathrm{out}} e^{-\beta\Delta E} \right]}
\end{equation}
\begin{equation}
    \mbox{acc}(o_\mathrm{in}\rightarrow n_\mathrm{out}) = \min{\left[1, \frac{p_\mathrm{bias}V_\mathrm{out}}{(1-p_\mathrm{bias})V_\mathrm{in}} e^{-\beta\Delta E} \right]}
\end{equation}
where subscripts denote the position of particle $i$ relative to particle $j$.

This method enhances the probability of removing or adding a particle from or to a cluster. However, AVBMC does not specifically facilitate cluster fusion, fission, or diffusion. To solve this problem, Wu suggested combining single-particle moves and cluster motions~\cite{wu1992electrostatic}, which we also examine below.

\subsection{Event-Chain Monte Carlo}
In order to avoid the high rejection rate of standard MMC in SALR models, we also consider an extended version of the rejection-free ECMC algorithm \cite{michel2014generalized,bernard2009event}, which breaks detailed balance but still satisfies (global) balance. In this scheme, a randomly selected particle $i$ moves in a preassigned direction with a succession of infinitesimal displacements, until it collides with a particle $j$ that then continues the motion through a process called lifting, which is reminiscent of momentum transfer. Here a factorized Monte Carlo filter is applied $p^\mathrm{fact}(o \rightarrow n) = \prod\limits_{i,j}\min{[1, \exp{(-\Delta E_{ij})}]}$, which is parallel to the Metropolis acceptance probability, $\mathrm{acc}(o \rightarrow n) = \min{[1, \exp{(-\sum\limits_{i,j}\Delta E_{ij})}]}$. Lifting happens if the particle move is rejected, thus efficiently overcoming spatial traps. 

More specifically, the displacement of $i$ before lifting $j$, $s_{ij}$, is obtained from 
\begin{equation}
\label{eq:ecmc}
    E_{ij}^* = \int_0^{E_{ij}^*}[dE_{ij}]^+ = \int_0^{s_{ij}}\left[\frac{\partial E_{ij}(\mathbf{r}_j - \mathbf{r}_i - s\mathbf{e}_k)}{\partial s}\right]^+ ds
\end{equation}
where $[dE_{ij}]^+$ denotes the energy increase for an infinitesimal displacement, and $E_{ij}^*$ is the admissible energy increase, such that $E_{ij}^*=-\ln\gamma_{ij}$ with $\gamma_{ij}$ a number taken uniformly at random over the interval $[0,1)$~\cite{michel2014generalized}. For efficiency, convenience, and without loss of generality\cite{michel2014generalized}, we take $\mathbf{e}_\mathbf{k}$ with $\mathbf{k}\in\{\hat{\mathbf{x}},\hat{\mathbf{y}},\hat{\mathbf{z}}\}$. After calculating the set $\{\lvert s_{ij}\rvert\}$, for $\forall j \not= i$, its minimal value is taken as the actual displacement of $i$, thus identifying the position and the collision partner of the next lifting. When the total displacement of an \emph{event chain} equates the preset chain displacement $l$, the last moving particle stops. 

Incidentally, this scheme also yields an unbiased estimator of the pressure equation of state
\begin{equation}
\begin{split}
    \frac{\beta P}{\rho} &= \left< \frac{x_\mathrm{final}-x_\mathrm{initial}}{l} \right>_\mathrm{chains} \\
    &= \left< \frac{l+\sum\limits_\mathrm{lifts}(x_j-x_i)}{l} \right>_\mathrm{chains}
\end{split}
\end{equation}
where $\langle\,\cdot\,\rangle$ is the average over event chains, and $\sum\limits_\mathrm{lifts}(x_j-x_i)$ is the excess displacement from lifting. In other words, the numerator is the difference between the positions of the first and the last particle in the event chain, and the denominator is sum of individual particle displacements, $l$, by construction. In practice, we find this on-the-fly estimator to more efficiently suppress short-time fluctuations of pressure than the traditional virial theorem route. 

\section{Time Correlation functions}
\label{sec:observables}
In order to assess the efficiency the advanced sampling algorithms described in Sec.~\ref{sec:MCmethods}, we need to quantify how relevant structural features decorrelate. Because the cluster, percolated and void fluids of disordered microphases have markedly distinct structures, different features then ought to be assessed as well. In this section, we present the combination of local and global observables used to evaluate the sampling dynamics in these different regimes.

\subsection{Cluster Fluid}
In the (particle) cluster regime, we monitor the local reorganization of particles through updates in cluster composition, and the evolution of the overall system structure by tracking cluster displacements.

The \emph{particle colocation} correlation function probes particle exchanges between clusters and their environments as well as cluster reorganization, as has been used in other micelle and cluster-forming systems~\cite{johnston2016toward,jadrich2015equilibrium}. More specifically, the colocation function for particle pair $\{i,j\}$
\begin{equation}
\theta_{ij} = 
    \begin{cases}
        1 & \textrm{if $i$ and $j$ are in the same cluster}\\
        0 & \textrm{if $i$ and $j$ are in different clusters}
    \end{cases}
\end{equation}
is used to evaluate the probability that a pair of particles initially, at time $t_0$, in the same cluster remains in a same cluster a time $t$ later -- even though cluster indexes might have changed in the mean time --
\begin{equation}
\label{eq:cpair}
    C_\mathrm{pair}(t) 
    = \frac{\sum\limits_{i,j}(\langle\theta_{ij}(t_0)\theta_{ij}(t_0+t)\rangle - \langle\theta_{ij}\rangle^2)}{\sum\limits_{i,j}(\langle\theta_{ij}^2\rangle - \langle\theta_{ij}\rangle^2)}.
\end{equation}
Note that normalizing the averages $\langle\cdots\rangle$ by the instantaneous cluster size $(s_{ij}(t_0)s_{ij}(t_0+t))^{1/2}$ accounts for cluster size fluctuations.

The \emph{cluster translation} correlation function monitors cluster displacements, which are typically much slower than their single-particle counterparts. We thus define an overlap function that monitors the cluster center of mass position $r_i$
\begin{equation}\label{eq:ccluster}
    C_\mathrm{clus}(t) = \frac{1}{N_\alpha}\sum_{\alpha}w(a\sigma-\min_{\beta}(\lvert \mathbf{r_\alpha}(t_0) - \mathbf{r_\beta}(t_0+t)\rvert)),
\end{equation}
where $\alpha$ and $\beta$ are cluster indexes at times $t_0$ and $t_0+t$, respectively, and $N_\alpha$ is the number of clusters at $t_0$. The term $r=\min{(\lvert \mathbf{r_\alpha}(t_0) - \mathbf{r_\beta}(t_0+t)\rvert)}$ captures the distance between centers of mass of the initial cluster $\alpha$ and the nearest cluster $\beta$ after time $t$. Without loss of generality, the overlap function is taken to be a step function, 
\begin{equation}\label{eq:wr}
    w(a\sigma-r) = \left\{ \begin{array}{ll}
              1 & \mbox{for $0 \leq r \leq a\sigma$};\\
              0 & \mbox{otherwise}.\end{array} \right.
\end{equation}
where $a\sigma$ is a constant on the scale of the typical cluster radius. 

In the cluster fluid regime, aggregates can either be classified as stable clusters or as metastable subclusters~\cite{johnston2016toward}, depending on their sizes. The former are the abundant species that form beyond the ccd; the latter are intermediate between clusters and single particles. (In practice, the local minimum of cluster size distribution is taken as separation threshold, which 
is roughly $n = 7$ (Sec.~\ref{sec:discussion}), but the analysis is relatively insensitive to the precise threshold chosen.) Because subclusters are small and fairly unstable, they diffuse and reorganize faster than stable clusters. In order to capture the genuinely slow structural relaxation time, we thus exclude subclusters from this correlation function. For the specific model considered, the remaining stable clusters then typically have either one or two particle shells, hence we set $a=2$.

For the function in Eq.~\eqref{eq:ccluster}, at time $t=t_0$, and hence $C_\mathrm{clus}(0)=1$. The long-time limit $C_\mathrm{clus}(t\rightarrow\infty)$ goes to a small constant proportional to the number density of clusters and the overlap volume. This trivial contribution is subtracted before evaluating the correlation time.

\subsection{Percolated Fluid}
As density increases, clusters elongate and eventually percolate. This makes cluster-based structural observables essentially irrelevant, because most particles are then part of the same cluster, and cluster displacements are but trivial translations of the system center of mass. In order to evaluate the system relaxation, different dynamical correlations are thus needed. 

The \emph{bonding} correlation function, which is commonly used in studies of gelation \cite{del2008network,miller2009dynamical}, is akin to $C_\mathrm{pair}$ for clusters in Eq.~\eqref{eq:cpair}. Rather than attachment to a cluster, it follows updates to more immediate particle connections, 
\begin{equation}\label{eq:p_bond}
    C_\mathrm{bond}(t) 
    = \frac{\sum\limits_{i,j}(\langle n_{ij}(t_0) n_{ij}(t_0+t)\rangle - \langle n_{ij}\rangle^2)}{\sum\limits_{i,j}(\langle n_{ij}^2\rangle - \langle n_{ij}\rangle^2)},
\end{equation}
where $n_{ij}$ characterizes the bonding between particles $i$ and $j$,
\begin{equation}
    n_{ij} = \left\{ \begin{array}{ll}
    1 & \mbox{if $i$ and $j$ are bonded};\\
    0 & \mbox{otherwise}.\end{array} \right.
\end{equation}\\

The \emph{network translation} correlation function is similarly an extension of Eq.~\eqref{eq:ccluster}. It tracks displacements of pairs of particles within the percolated component, 
\begin{equation}\label{eq:p_particle}
    C_\mathrm{ntwk}(t) = \frac{1}{N}\sum\limits_{i}w(b\sigma-\min_{j}(\lvert \mathbf{r}_i(t_0) - \mathbf{r}_j(t_0+t)\rvert)),
\end{equation}
where $w(b\sigma-r)$ is given by Eq.~\eqref{eq:wr}, with $b=1$ set to approximate the extent bonded particle motion. The minimization, which provides the distance between the particle $i$ at time $t_0$ and the closest particle $j$ at time $t_0+t$, treats particles indistinguishably. As a result, this correlation function traces global structural changes. 

\subsection{Void Cluster Fluid}
At even higher densities, distinct aggregates of void space (void clusters) evolve within a near continuum of particles. The decorrelation of the void structure is then an important dynamical consideration \cite{zhuang2017communication}. Tracking the decorrelation of particle-based structures alone may indeed miss out on the persistence of the voids, especially as their relative contribution diminishes.

In a lattice-gas SALR model, a clear duality exists between voids and particles (see, \emph{e.g.}, Ref.~[\onlinecite{tarzia2021}]), but in continuous space clusters of void space are somewhat more ambiguously defined. Various approaches previously considered, including Voronoi tessellation \cite{rintoul2000precise}, soft particle substitution \cite{lindquist2016assembly}, and small particle filling \cite{zhuang2017communication}, have some merits. While these schemes are well defined geometrically, they do not lead to an obvious microscopic measure of structural decorrelation. We thus here consider a scheme based on gridding space into small spherical cells of diameter $\sim0.5\sigma$ placed on a simple cubic lattice. The structure is chosen over standard cubic cells for calculational convenience and simulation efficiency; the scale is taken as slightly larger than the trivial triangular gap between close-packed, coplanar particles of diameter $1.5\sigma$, \emph{i.e.}, the extent of the interparticle attraction. A cell is deemed empty if it is out of the attraction range of any particle. (This purely geometric definition does not correct for rare fluctuations that open up cavities at high temperatures.) The generalization to dynamics is then straightforward. From the occupancy of cell $i$
\begin{equation}
    v_{i} = \begin{cases}
    1 & \textrm{if $i$ is empty}\\
    0 & \textrm{otherwise}
    \end{cases},
\end{equation}
one can indeed define the \emph{void} correlation function
\begin{equation}
\label{eq:v_void}
    C_\mathrm{void}(t) 
    = \frac{\sum\limits_{i}(\langle v_{i}(t_0) v_{i}(t_0+t)\rangle - \langle v_{i}\rangle^2)}{\sum\limits_{i}(\langle v_{i}^2\rangle - \langle v_{i}\rangle^2)}
\end{equation}
which measures the persistence of a void cell with time. 
This grid-based approach, however, is less accurate than other schemes. In particular, partially empty cells are not counted and small voids can be missed. The void volume it identifies is therefore smaller than the actual one.

The second dynamics considered in this regime is that of the particles. Because they form a nearly homogeneous component, standard \emph{self-overlap} correlation functions from liquid state studies (see, \emph{e.g.}, Refs.~\onlinecite{kob1995testing,yamamoto1998heterogeneous,pedersen2010geometry}) are then appropriate. We thus consider
\begin{equation}
\label{eq:v_self}
    C_\mathrm{self}(t) = \frac{1}{N}\sum\limits_{i}w(b\sigma-\lvert \mathbf{r_i}(t_0) - \mathbf{r_i}(t_0+t)\rvert).
\end{equation}
where $w(b\sigma - r)$ is as in Eq.~\eqref{eq:wr}, with $r$ of a single-particle displacement. 

\subsection{Non-Exponential Relaxations}
\begin{figure}
\centering
\includegraphics[width=8.5cm,trim={1cm 0cm 0 1cm},clip]{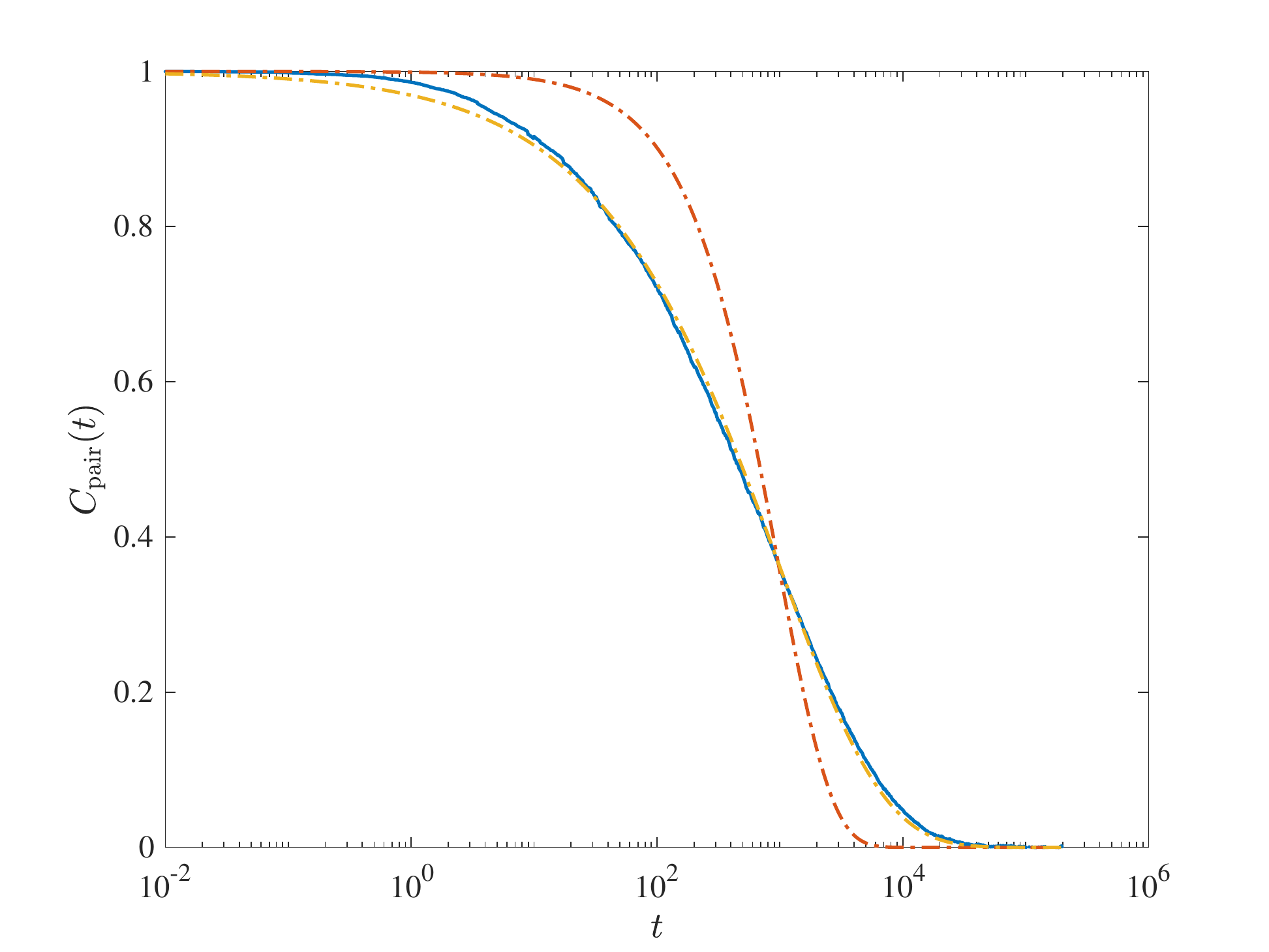}
\caption{Time decorrelation of the particle colocation function from Eq.~\eqref{eq:cpair} by AVBMC sampling (blue) at $T = 0.7$, $\rho = 0.05$. A stretched exponential form (yellow) with $\tau = 97$, $\zeta = 0.50$ clearly better captures the function than an exponential form (red) with a same $\tau$.} 
\label{fig:fitting}
\end{figure}

In order to extract characteristic relaxation times $\tau$  from these correlation functions, an exponential fitting form does not suffice. As shown in Fig.~\ref{fig:fitting}, the decay of most of these relaxation processes is better captured by a stretched exponential form \cite{klafter1986relationship,ediger2000spatially,bouchaud2008anomalous},
\begin{equation}
\label{eq:stretchedexp}
C(t) \propto e^{-(t/\tau)^\zeta}, \ \ \zeta<1.
\end{equation}
with stretching exponent $\zeta$. Stretched exponentials, which are commonly used to describe the dynamics of dense disordered systems \cite{ediger2000spatially,larson1999structure}, can suggest both parallel or serial relaxation processes \cite{palmer1984models,klafter1986relationship}. The former refer to independent but simultaneous relaxation of different degrees of freedom, and the latter refer to sequential decorrelation steps. In cluster fluids, for instance, clusters of various sizes and shapes coexist, each with different relative stability and thus associated lifetime. Combining their relaxation processes results in $C(t)$ decaying non-exponentially \cite{hoy2015structure,gelbart2012micelles}. Void cluster fluids, as duals of particle cluster fluids, experience similar heterogeneous relaxation. 
The structure of percolated fluids is also heterogeneous. Different structures likely evolve over different relaxation timescales. In addition, a molecular dynamics study suggests that multiple relaxation steps, including short-time collisions and long-time (activated) bond breaking, are then at play\cite{del2008network}. In any event, both effects result in strongly heterogeneous dynamics.  
Ample evidence thus motivates using Eq.~\eqref{eq:stretchedexp} as fitting form, and $\zeta$ to vary from one regime to another upon changing the sampling algorithms. 

\begin{figure}[t!]
\centering
\subfigure{
\begin{minipage}{\columnwidth}
\centering
\includegraphics[width=8.5cm,trim={0.8cm 0.2cm 0cm 0cm},clip]{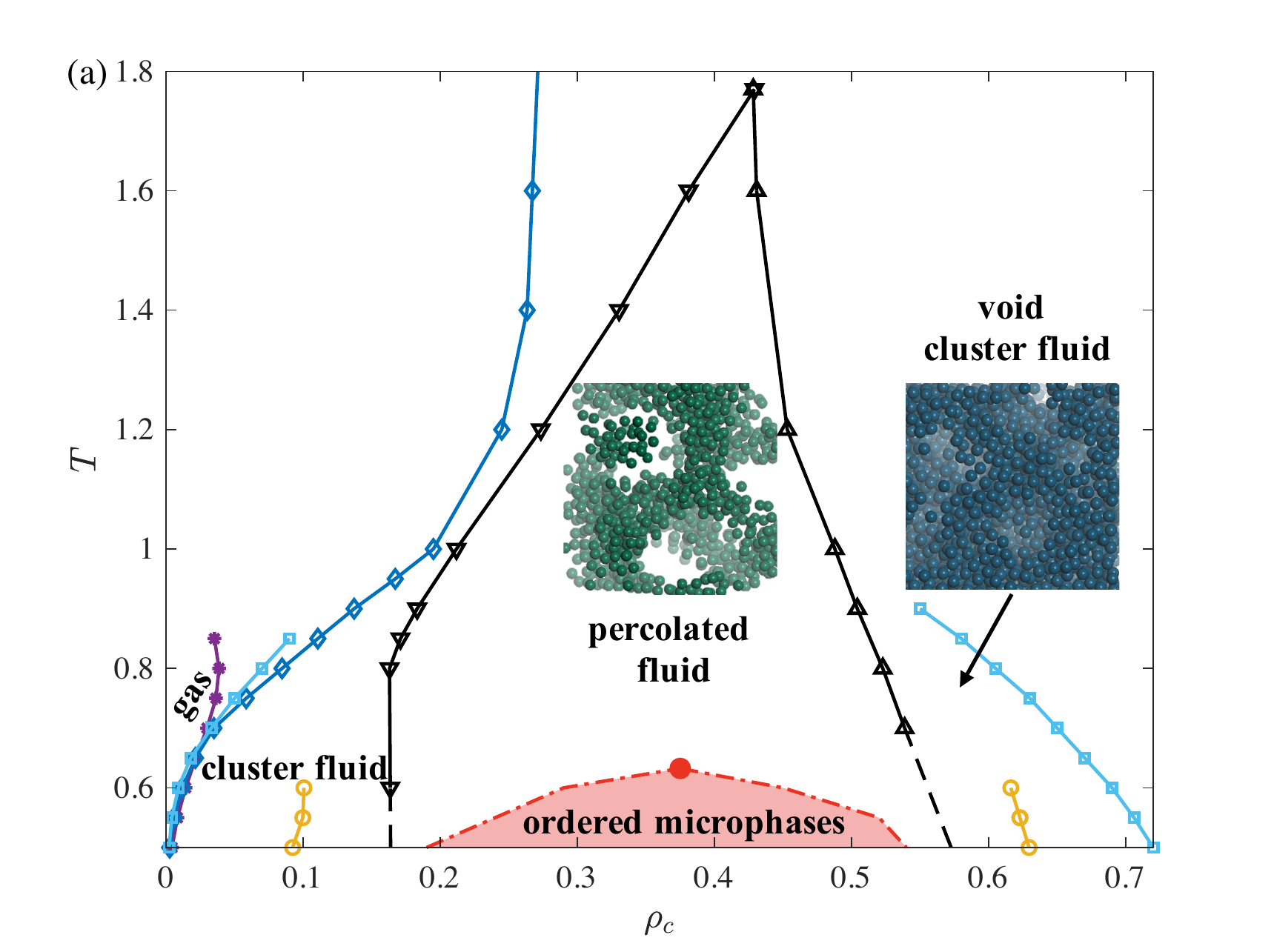}
\end{minipage}
}
\subfigure{
\begin{minipage}{\columnwidth}
\centering
\includegraphics[width=8.5cm,trim={0.8cm 0.2cm 0cm 0cm},clip]{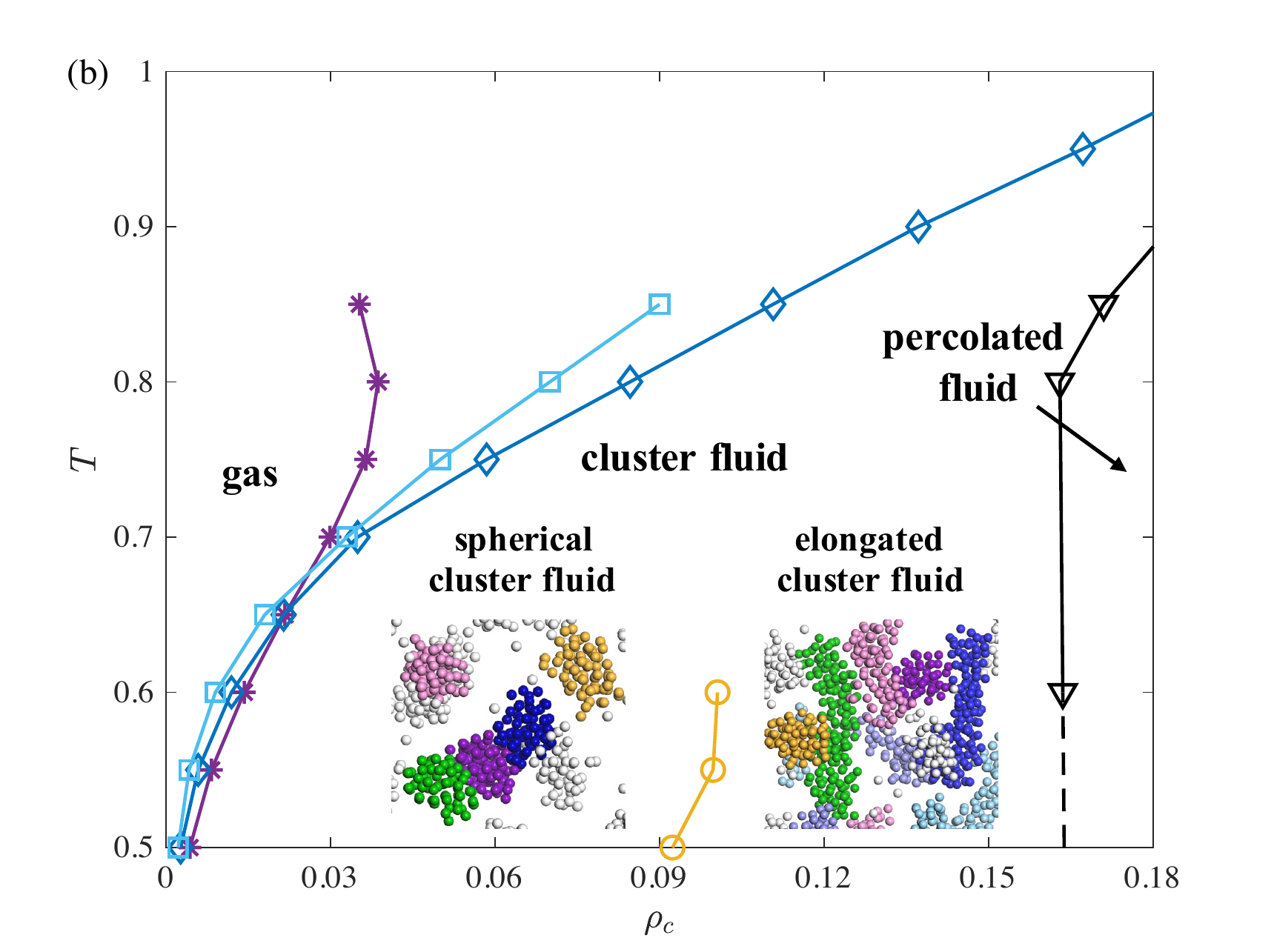}
\end{minipage}
}
\caption{$T$-$\rho$ phase diagram of the SALR model specified in Sec.~\ref{sec:model}, highlighting the various structural regimes of disordered microphases. The ccd estimates from $h(\rho)$ (purple stars, as in Fig.~\ref{fig:hrho}) and from $C_V$ (blue diamonds) coincide at low $T$, but eventually separate. The former terminates at roughly the same $T$ as that of the clustering onset from the cluster size distribution $s\Pi(s)$ (cyan cubes), while the latter tracks the clustering density from $s\Pi(s)$ but without non-trivial end point. At lower $T$, the particle and void clustering regimes subdivide between spherical and elongated clusters (yellow circles), as obtained from the cluster anisotropy distribution. The particle physical (black downward triangles) and void geometrical (black upward triangles) percolation threshold lines delineate the percolated fluid (bicontinuous gel) regime. The dome of ordered microphases (red dot-dashed line) is estimated from self-assembly simulations, and is here only roughly sketched. Otherwise, errors are comparable to the symbol sizes.}
\label{fig:phase}
\end{figure}

\section{Disordered Microphases}
\label{sec:discussion}

In this section, we provide a thorough investigation of the various structural regimes of the high-temperature fluid phase in which equilibrium disordered microphases are observed. It should be emphasized that the (void) cluster fluid and (void) percolated fluid regimes are not proper thermodynamic phases as they are only separated by crossovers. The positions of the resulting separation lines thus depend on the choice of observables. Although our choices are motivated in part by the aim of assessing the algorithmic efficiency of various MC sampling schemes in Sec.~\ref{sec:dynamics}, they are not unique.

In the following we present specific observables used to detect particle and void clustering, cluster asphericity, percolation and the ODT. In all cases, a minimum of 20 replicate simulations are run for at least $10^5$ and up to $10^6$ MC moves at each state point in the constant $NVT$ ensemble. The sampling duration depends on the regime considered and on the sampling algorithm used, but in each case proper equilibration and then sufficient sampling are achieved (reaching at least $10\tau$ based on the dynamical results of Sec.~\ref{sec:dynamics}). In general, low-temperature conditions and the vicinity of the ODT require more system replicas and longer sampling to achieve proper averaging.

\subsection{Order-disorder Transition (ODT)}
The ODT that replaces the gas-liquid critical point in SALR models with sufficiently strong repulsion is reckoned to be weakly first-order for isotropic interactions \cite{brazovskiui1996phase,seul1995domain,bates1990block}. This transition identifies the highest temperature at which ordered periodic microphases--which for sufficiently symmetric interactions are lamellar--can be observed. Zhuang \emph{et al}.~\cite{zhuang2016equilibriumJPC} have estimated the ODT for a related SALR model based on the melting of that order. Because we don't have here access to equilibrated configurations at $T<T_{\mathrm{ODT}}$, we instead estimate that transition using the local maximum of the heat capacity. Although admittedly somewhat imprecise, this estimate nevertheless aptly serves the purpose of bounding the $T-\rho$ dome of ordered microphases, so that disordered microphases can be studied without interference.

The weakly first-order nature of the ODT suggests that the heat capacity peak should first exhibit a critical-like scaling with system size, until a first-order like discontinuity is reached. Most of this phenomenology is, however, beyond the reach of our current simulation scheme and computational resources. As a result, our ODT estimate is but an upper bound to the thermodynamic transition. The large error bar in Fig.~\ref{fig:phase} reflects the crudeness of this approach. More accurate ODT will hopefully be obtained in the future from improved sampling algorithms. For now, it suffices to note that for $T \geq 0.65$ all conditions can confidently be deemed to lie in the equilibrium high-temperature disordered phase. We here mostly consider systems above that temperature. Only at densities far below or far above that of the ODT, do we consider systems at lower temperatures. In these regimes, microphase ordering emerges at even lower $T$, and thus our simulations remain safely distant from it.

\subsection{Particle Clustering}
For sufficiently low temperatures, increasing density leads to a sudden onset of clustering. As commonly argued, this critical cluster density (ccd) of SALR systems is akin to the critical micelle concentration (cmc) of surfactants \cite{presto1948some,johnston2016toward,santos2016determination,hu2018clustering,pekalski2019self}. Two main thermodynamic observables have been suggested to detect this transformation. First, following the standard micellization procedure~\cite{pekalski2019self}, deviations from the ideal gas equation of state (or $h(\rho)$ from Eq.~\eqref{eq:h_rho}) are often considered (see Sec.~\ref{sec:model}). The microscopic motivation is that in the aggregation process--from a gas of particles to a gas of clusters--the number density of independent species goes down markedly. As a result, the growth rate of pressure changes at the onset of clustering  captured by $h(\rho)$. Second, the heat capacity detects the instability of the homogeneous fluid upon the rapid increase of thermal excitation that accompanies cluster formation~\cite{frantz1995magic,schwanzer2016two,pekalski2019self,imperio2006microphase}. In two-dimensional studies of lattice and off-lattice models \cite{pekalski2019self,schwanzer2016two}, $C_V$ was found to predict an onset of clustering consistent with that given by direct cluster detection schemes. Because $C_V$ captures structural rearrangements indiscriminately and not specifically clustering, however, in more complex systems (such as three-dimensional off-lattice models) other excitations could obfuscate this identification. 

At higher $T$, the local minimum of $h(\rho)$, which follows from the instability of the homogeneous fluid to clustering at low $T$ (see Appendix~\ref{sec:appendix}), fails to detect any structural inhomogeneity. (Interestingly, percolation also goes unnoticed.) By contrast, the heat capacity peak shifts to higher $\rho$ and persists even beyond the percolation line. This last regime is clearly unrelated to clustering and likely stems from degrees of freedom other than cluster (dis)assembly then becoming thermally accessible. In particular, the next dominant contribution to $C_V$ is facile cluster scission and reformation, as observed. (Rotations or combinations of vibration and rotation have also been proposed \cite{touloukian1970thermophysical,frantz1995magic}, but are not observed in our simulations.) The peak in $C_V$ weakens and broadens as $T$ increases, and is expected to eventually vanish, because the pure hard sphere limit exhibits no such peak. The question, however, remains. In the cluster regime, which observable better tracks the actual clustering process? We get back to this point after having examined the cluster size distribution more completely.

\begin{figure}
\centering
\subfigure{
\begin{minipage}[t]{0.99\linewidth}
\centering
\includegraphics[width=8.5cm,trim={0.8cm 1.2cm 1cm 1cm},clip]{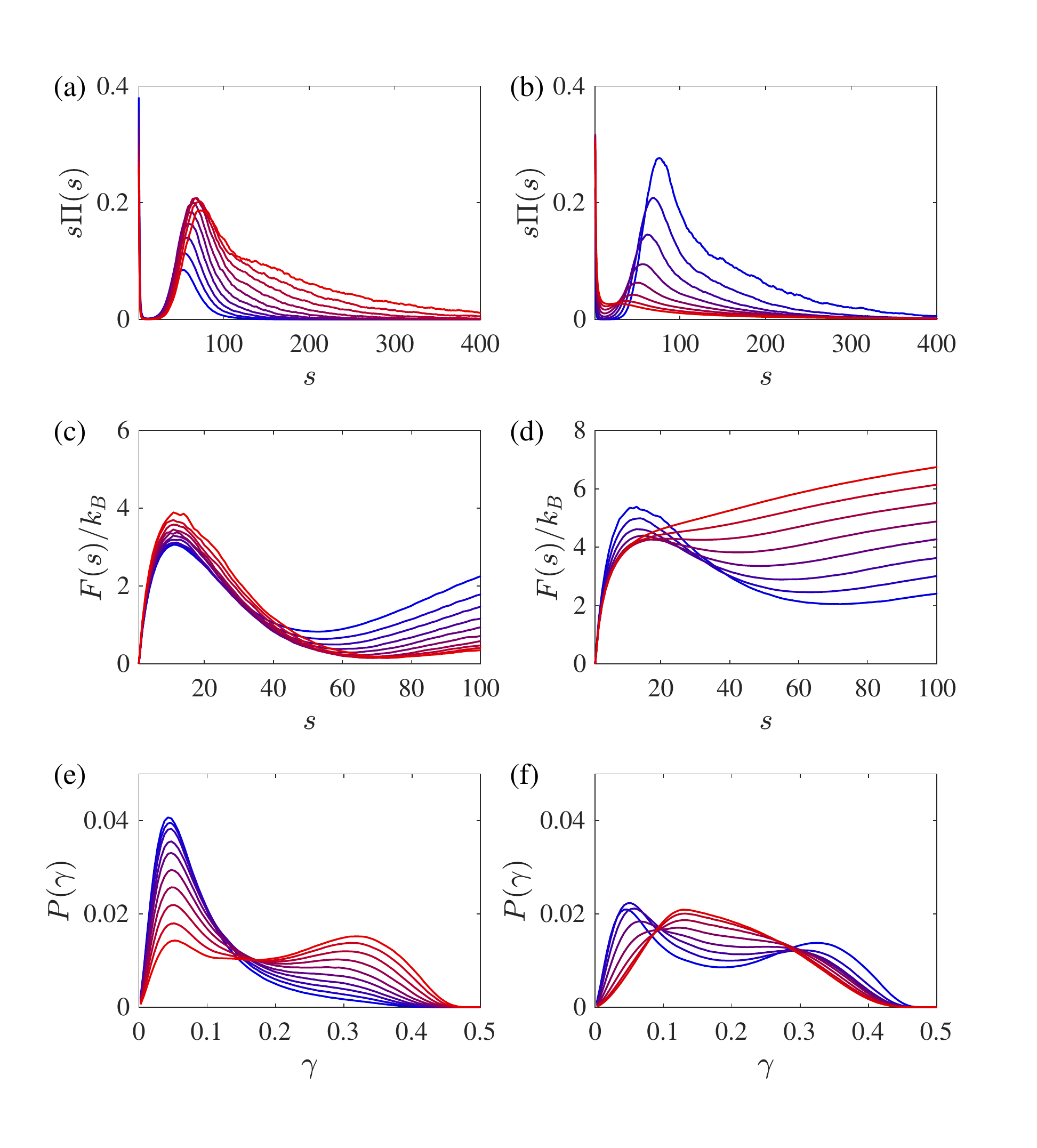}
\end{minipage}
}
\caption{Evolution of (a,b) the cluster size distribution, (c,d) the free energy of cluster formation, and (e,f) the cluster anisotropy distribution for (a,c,d) $T = 0.55$ with density $\rho = 0.03$-$0.12$ in 0.01 increments (from blue to red) and for (b,d,f) $\rho=0.1$ with $T = 0.5$-$0.85$ in 0.05 increments (from blue to red). Note that the equilibrium monomer free energy is arbitrarily set to zero, and should thus not be confused with the free energy of a gas-state particle.}
\label{fig:cdf}
\end{figure}

Particle clusters in the SALR model studied here (and in others~\cite{tarzia2021}) are structurally a lot richer than typical spherical micelles, which tend to have fairly narrowly distributed features. Consider, in particular, the distribution $s\Pi(s)$ of clusters of size $s$~\cite{hu2018clustering}, defined using the attraction range $\lambda_0\sigma$ as a geometric bonding criterion. Beyond the ccd, two peaks can be found: one for monomers and the other for clusters. As expected, the monomer peak is narrow, but the cluster peak is quite broad (Fig.~\ref{fig:cdf}(a-b)). As density increases, in particular, we find that a significant fraction of particles are part of clusters whose sizes are multiple times that of the typical cluster. (Morphological changes also accompany this effect, as further discussed in Sec.~\ref{sec:anisotropic}.) Transforming $\Pi(s)$ into an effective free energy~\cite{mysona2019mechanism,mysona2019simulation}
\begin{equation}
    \beta F(s)\propto-\ln [\Pi(s)],
\end{equation}
confirms that the cluster free energy profile is quite flat, and hence that large size fluctuations are thermally accessible (Fig.~\ref{fig:cdf}(c-d)). The cluster peak is nevertheless well separated from the single particle peak. A marked free energy barrier separates the two wells, hence clearly defining the sub-cluster metastability threshold \cite{santos2017thermodynamic} (see Sec.~\ref{sec:observables}). 
In order to understand physical trends in this regime, it is helpful to recall that this effective free energy profile results from the competition between the entropy of dispersion and the enthalpy of aggregation. The former favors monomers, while the latter favors clusters \cite{gelbart2012micelles}.

%figure: thermodynamic properties

As temperature increases, the cluster peak in $s\Pi(s)$ steadily vanishes while the monomer peak gains mass. As expected, entropy then dominates over enthalpy. The shrinking distinction between the two peaks is reminiscent of a second-order phase transition, with the barrier in $F(s)$ vanishing roughly where the ccd from $h(\rho)$ disappears ($T = 0.85$ as shown in Fig.~\ref{fig:phase}). However, the estimates of ccd from $s\Pi(s)$ are closer to those from $C_V$, as for the lattice model considered in Ref.~[\onlinecite{tarzia2021}].
This analogy suggests a thermodynamic definition of clusters, with the cluster peak in $s\Pi(s)$ vanishing approximately around the $C_V$ peak. This is not, however, the case here, probably as a result from the richer set of degrees of freedoms here contributing to the heat capacity.

As $\rho$ increases beyond the ccd, the cluster peak gains mass, broadens, and shifts to larger sizes, while the monomer peak simply shrinks (Fig.~\ref{fig:cdf}). (The anomalous decrease in cluster peak height above $\rho=0.1$ comes from percolating clusters absorbing a large fraction of particles.) In $F(s)$, correspondingly, the local free energy minimum of clusters decreases, flattens and also shift to larger sizes, while the barrier to clustering becomes more pronounced, reaching more than $5k_B T$. Cluster size equilibration through single-particle addition and removal is then expected to get markedly less efficient, which \emph{a posteriori} motivates the consideration of more collective sampling mechanisms. In any case, this clustering trend is not straightforwardly explained by the enthalpy-entropy framework, and is instead considered in the following section (Sec.~\ref{sec:anisotropic}).

\begin{figure}
\centering
\includegraphics[width=8.5cm,trim={0.2cm 0.5cm 0cm 0cm},clip]{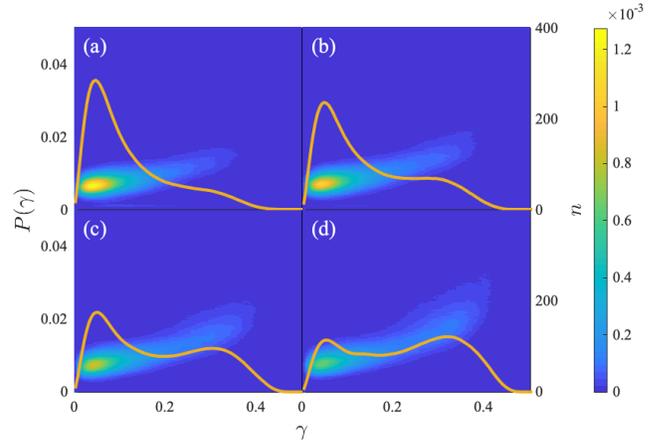}
\caption{Cluster size (right $y$ axis) and shape ($x$ axis) joint distribution, $P(\gamma,n)$ at $T = 0.55$ for $\rho = $ (a) $0.06$, (b) $0.08$, (c) $0.1$, and (d) $0.12$. The heat map encodes the probability density. The cluster anisotropy distribution, $P(\gamma)$ (yellow curves) (left $y$ axis) from Fig.~\ref{fig:cdf} is given as reference.}
\label{fig:cadf_N}
\end{figure}

\subsection{Anisotropic Clustering}
\label{sec:anisotropic}
As density increases, clusters not only grow but become more aspherical. In order to quantify this shape evolution, we employ a common apshericity parameter (see, \emph{e.g.}, Ref.~\onlinecite{fusco2013crystallization})
\begin{equation}
    \gamma = \frac{(I_1 - I_2)^2 + (I_1 - I_3)^2 + (I_2 - I_3)^2}{2(I_1^2 + I_2^2 + I_3^2)}
\end{equation}
where $I_i$ denotes the $i$th moment of inertia of a given cluster. For an ideal spherical cluster, $I_1 = I_2 = I_3$ and hence $\gamma = 0$. Aspherical structures have $\gamma>0$. In particular, elongated, rod-like clusters have $\gamma \approx 1/2$; flattened, plate-like one have $\gamma \approx 1/6$. 

The asphericity parameter is computed for each cluster, excluding percolated and sub-clusters for which this quantity is not physically significant. The resulting joint probability distribution, $P(\gamma, n)$, captures the fraction of clusters with a specific anisotropy and size. The normalized probability of a particle belonging to a cluster with a specific $\gamma$ is then $P(\gamma) = \sum\limits_{n} nP(\gamma, n)$ (Fig.~\ref{fig:cdf} (e-f)). 

This distribution reveals that at low temperatures the cluster regime can itself be subdivided. At $T=0.55$, for example, for $\rho\lesssim0.08$ only nearly spherical clusters are found, while for $\rho\gtrsim0.08$ the fraction of particles forming elongated clusters markedly increases (Fig.~\ref{fig:cdf}(e)). In order to quantify the onset of asphericity at a fixed $T$, we identify a crossover density $\rho_a$, such that the two cluster asphericity peaks have equal mass (Fig.~\ref{fig:phase}).  This distinction vanishes -- similarly to that between monomers and clusters at higher temperatures (Fig.~\ref{fig:cdf}(b))-- and a broad continuum of cluster asphericity is then observed (Fig.~\ref{fig:cdf}(f)). This effect likely follows from the large shape fluctuations that spontaneously appear as clusters become steadily less stable (and free energy barriers vanish in Fig.~\ref{fig:cdf}(d)).

The relationship between cluster shapes and sizes, as given by $P(\gamma, n)$, is illustrated in Fig.~\ref{fig:cadf_N}.  As density increases, small clusters remain fairly spherical, but larger ones elongate. In order to explain this growth in asphericity, the micelle analogy is helpful. Recall that the radius of a globular micelle is controlled by the length of the surfactants it contains~\cite{gelbart2012micelles}. From the attraction range used for Eq.~\eqref{eq:hamiltonian}, spherical clusters can similarly fully stabilize only one or two particle shells. Beyond this size, repulsion contributes significantly, which sets a natural size cutoff to the cluster size. However, as the overall system density increases cluster-cluster repulsion eventually disfavors simply forming additional small spherical clusters. Cluster elongation is therefore inevitable, and once asphericity emerges further extension comes at only a modest energetic cost. Percolation naturally follows at higher $\rho$. This elongation also explains the aforementioned trend in Fig.~\ref{fig:cdf}(a). As $\rho$ increases, the cluster peak in $s\Pi(s)$ shifts to larger sizes with greater size fluctuation as a result of spherical clusters being replaced by larger, elongated and easily fluctuating clusters.

\subsection{Percolation Lines}
\begin{figure}
\centering
\subfigure{
\begin{minipage}[t]{0.99\linewidth}
\centering
\includegraphics[width=8.5cm,trim={1cm 0.1cm 1.5cm 0cm},clip]{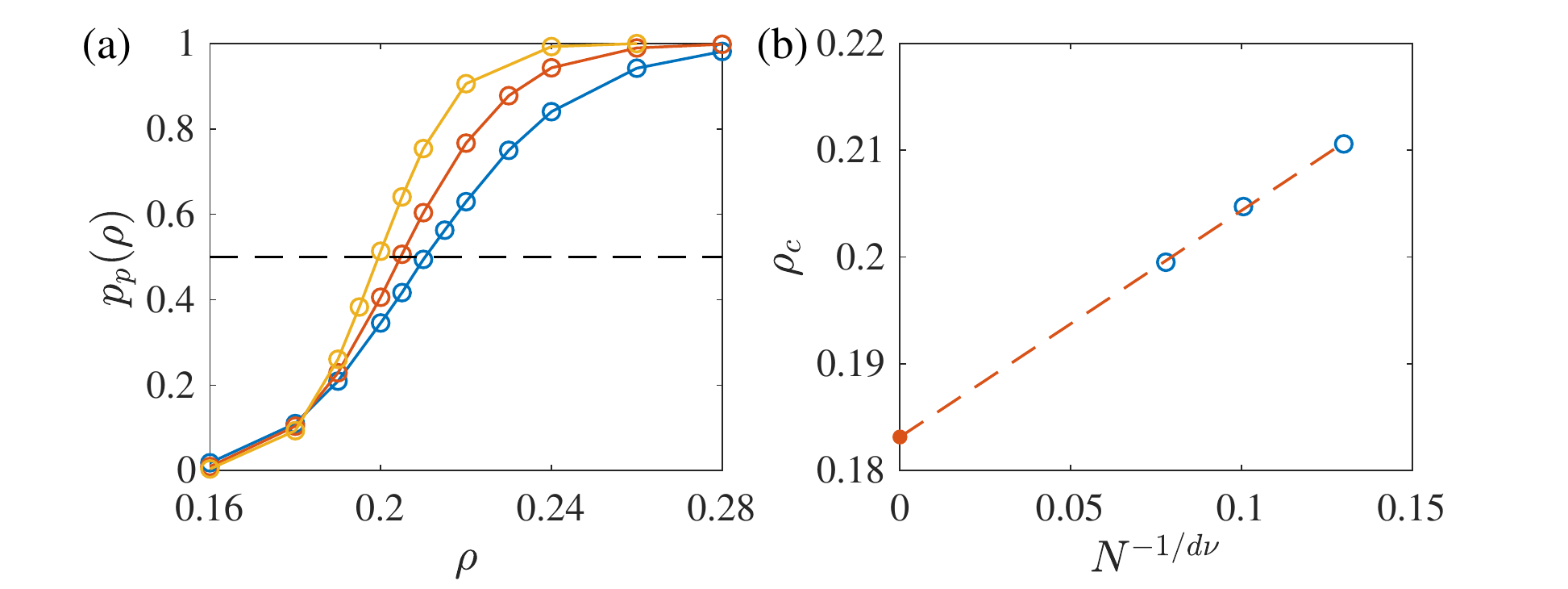}
\end{minipage}
}
\caption{Physical percolation threshold determination using the Coniglio-Klein bonding criterion, $p_\mathrm{B}=[1-\exp{(-\beta \varepsilon/2)}]$. (a) Density evolution of the percolation probability at $T = 0.9$ for $N = 250, 500, 1000$. (b) The probability midpoints are fitted with the critical scaling form of Eq.~\eqref{eq:percscaling} (red dashed line) to extract the thermodynamic threshold, here $\rho_c(0.9)= 0.183$.}
\label{fig:threshold}
\end{figure}

As hinted above, further increasing density gives rise to a percolated fluid. The onset of percolation is here considered using a physical criterion, instead of a purely geometric one, because as $T$ increases particle-particle attraction loses  thermal relevance. (In the cluster regime, $T$ is low enough that the two definitions are almost indistinguishable.) In the spirit of Coniglio and Klein~\cite{coniglio1980clusters}, we use a bond probability $p_\mathrm{B}=[1-\exp{(-\beta \varepsilon/2)}]$ to modulate the percolation probability. The percolation threshold is otherwise determined through standard scaling analysis~\cite{stauffer2018introduction}.
The density at which the percolation probability (fraction of configurations with a percolated cluster spanning in all three dimensions) attains $1/2$ in a finite-size system is then used in the critical scaling form to estimate the thermodynamic threshold, $\rho_c$
\begin{equation}
\label{eq:percscaling}
    \lvert \rho_{1/2}(N, T) - \rho_c(T) \rvert = N^{-1/d\nu},
\end{equation}
where the critical exponent $d\nu = 2.706$ for the  three-dimensional simple percolation universality class (Fig.~\ref{fig:threshold}). 
Note that at low $T$, percolation interferes with microphase ordering in small systems. (Finite-size systems artificially extend the dome of ordered microphases.) Only systems with $N\geq 1000$ are thus used for $T < 0.75$; above $T = 0.75$ systems with $N = 250, 500, 1000$ suffice.

As temperature increases, bonding weakens and particles disperse. Although both the percolation and the $C_V$ peak lines grow with density, they hardly track one another. Interestingly, a similar phenomenon has been reported for a patchy colloid model by Bianchi \emph{et al}.\cite{bianchi2008theoretical}, who associated this effect to bonding connectivity. At large connectivity, the $C_V$ peak lies below the percolation line, and vice versa. The former tracks bonding, while the latter tracks spanning structures, hence if connectivity is large, spanning structures form at low density, before particles are fully bonded. By contrast, the $C_V$ peak in Fig.~\ref{fig:phase} always lies above the percolation line, thus suggesting that the average thermodynamic connectivity is small in our system. Although the analogy is only qualitative, it confirms the general inconsistency of the two observables. 

Because of the approximate symmetry between particles and voids, the void percolation threshold is also instructive. Based on the definition of voids in Sec.~\ref{sec:observables}, two empty cells are deemed connected if they are immediate neighbors (26 in total). In contrast to the percolation analysis above, however, this definition is purely geometrical. No effective void \emph{bond energy} can be straightforwardly defined for off-lattice models (in contrast to lattice models, see, \emph{e.g.}, Ref.~\onlinecite{tarzia2021}). The resulting void percolation threshold should thus be close to the physically relevant one at low temperature, but get increasingly (thermally) irrelevant as temperature increases. From the results in Fig.~\ref{fig:phase}, one might speculate that the relevant void percolation line should more closely mirror that of particles, and hence the two lines should cross at lower $T$ than is here observed. In any event, the region contained below the two percolation lines roughly delimits the (bicontinuous) percolated (gel-like) fluid regime. At higher temperatures, the fluid is truly homogeneous for all (thermal) purposes. 

% percolation finite size scaling

\subsection{Void Clusters}
\begin{figure}[t!]
\centering
\subfigure{
\begin{minipage}{0.99\linewidth}
\centering
\includegraphics[width=8.5cm,trim={0.7cm 0.1cm 1.5cm 0cm},clip]{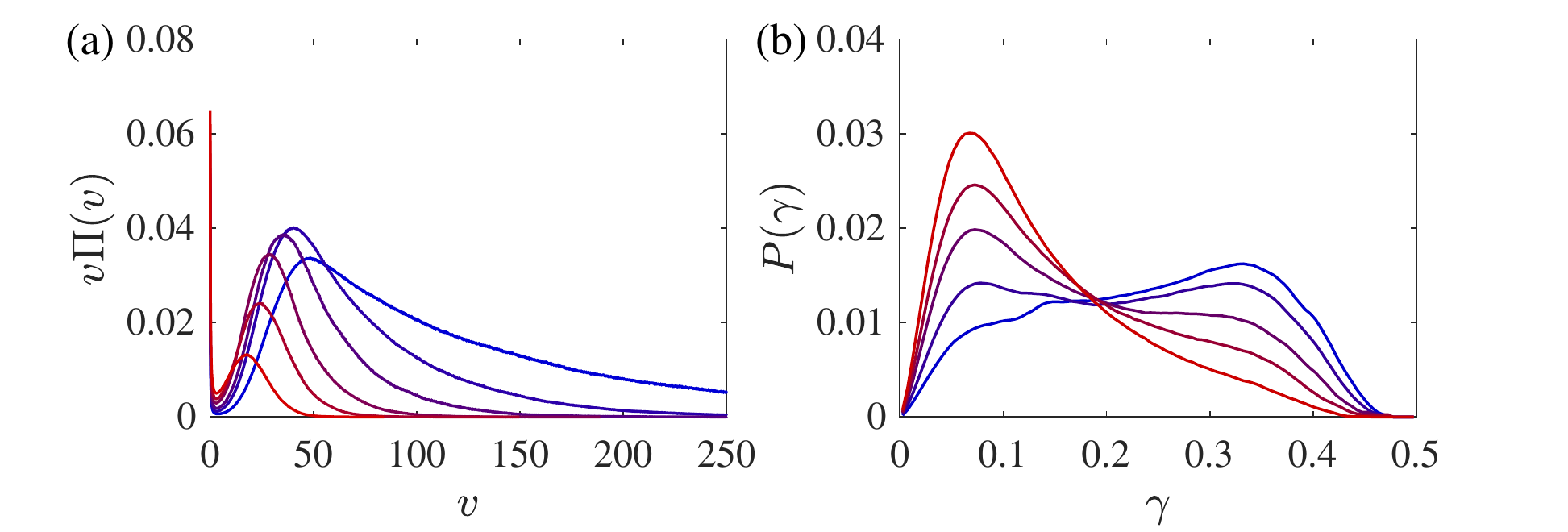}
\end{minipage}
}
\caption{Characterization of the void cluster fluid by (a) its void cluster size distribution, and (b) its void cluster anisotropy distribution at $T=0.6$ for $\rho = 0.58:0.02:0.66$ (from blue to red). Note that $N=5000$ is used to allow a sufficient number of void clusters to form. Each unit volume can roughly accommodate one particle (or $\sim 1.5$ particles in close-packing), so the results are comparable to Fig.~\ref{fig:cdf}. The approximate duality between voids and particles is thus confirmed at the geometric level.}
\label{fig:vdf}
\end{figure}

Like particles at low densities, void space clusters at high densities. We here consider the volume of voids (rather than the arbitrary number of void cells) to quantify void clustering. Figure~\ref{fig:vdf}(a) illustrates that two distinct peaks appear in the void volume distribution, which is reminiscent of the particle cluster size distribution (although a void monomer here cannot be similarly defined). As density decreases, the modal void volume shifts to larger values and its associated peak broadens, until void percolation is reached. The void cluster geometry in Fig.~\ref{fig:vdf}(b) suggests that these voids go from being nearly spherical to elongated as their density increases (particle density decreases), eventually leading to void space percolating, as discussed above. Therefore, similar crossovers for the onset of void clustering and of void anisotropy are determined in Fig.~\ref{fig:phase}.

These results demonstrate that the formal duality between particles and voids on lattices has a clear echo in off-lattice systems, in regard to the geometric structures. In a recent study of clustering using the Bethe approximation \cite{tarzia2021}, void aggregates are found to dominate thermodynamic properties at high densities, in the same way that particle aggregation dominates at low densities. The same is not true here. The equation of state is featureless in this regime and $C_V$ does not peak. This qualitative difference between on and off-lattice models probably originates from the abundance of degrees of freedom in the latter compared to the former. 

%figure: void properties

\section{Dynamics}
\label{sec:dynamics}
In this section, we consider the structural decorrelation dynamics in the cluster and percolated fluid regimes. (In the single-particle fluid regime, simple MMC sampling works fairly efficiently.) The algorithms described in Sec.~\ref{sec:MCmethods} are assessed using the local and global structurally-tailored correlation functions introduced in Sec.~\ref{sec:observables}. For a given sampling method, the slower of the two relaxation times is rate-limiting, and hence determines overall performance. Optimal MC algorithms can therefore be identified for the different disordered microphase morphologies.

\subsection{Cluster Fluid}
\label{sec:clusdyn}

\begin{figure}[htbp]
\centering
\subfigure{
\begin{minipage}[t]{0.48\textwidth}
\centering
\includegraphics[width=8.5cm,trim={0.8cm 0.2cm 0.4cm 0.8cm},clip]{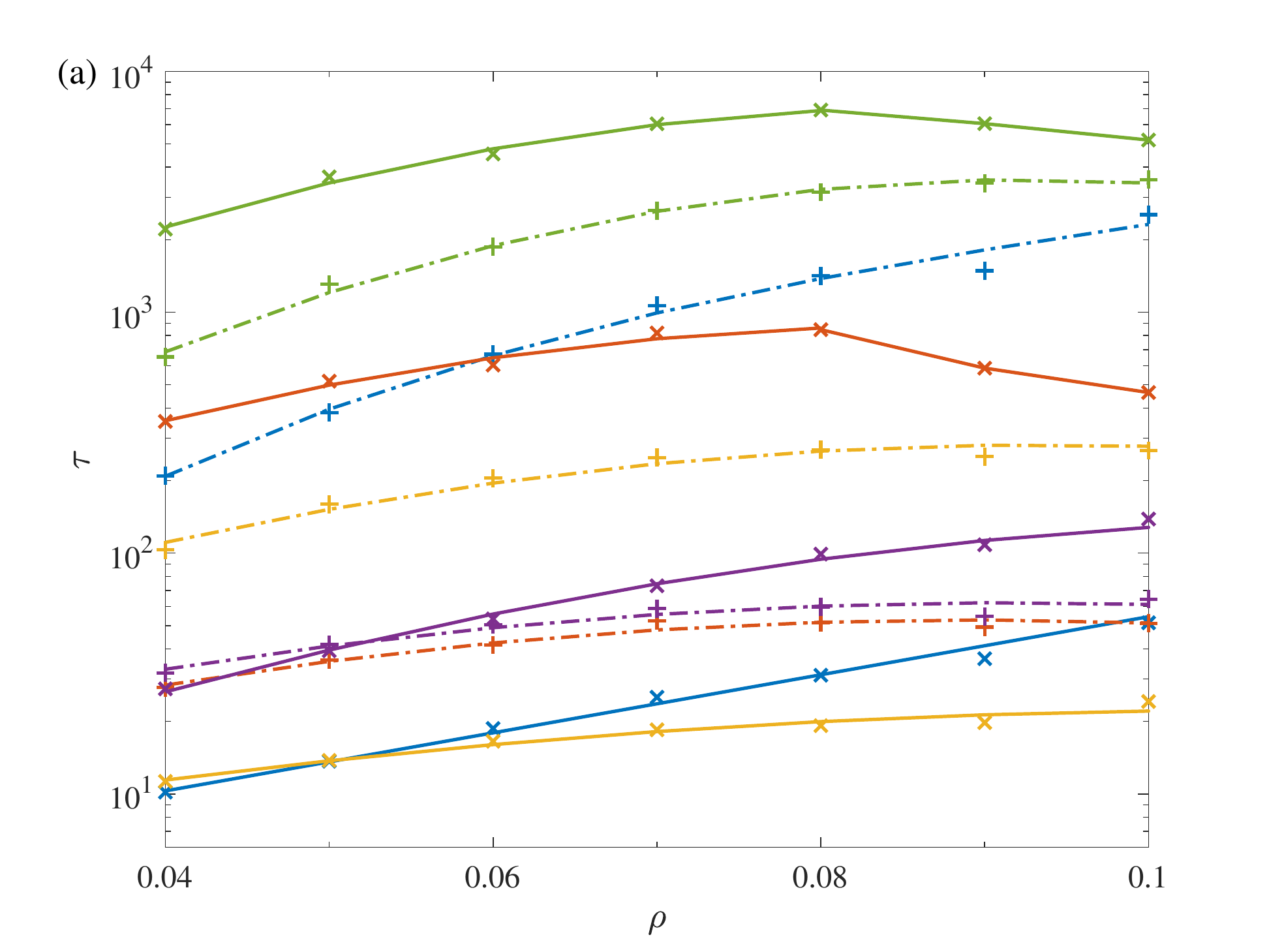}
\end{minipage}
}
\subfigure{
\begin{minipage}[tz]{0.48\textwidth}
\centering
\includegraphics[width=8.5cm,trim={0.8cm 0.2cm 0.4cm 0.8cm},clip]{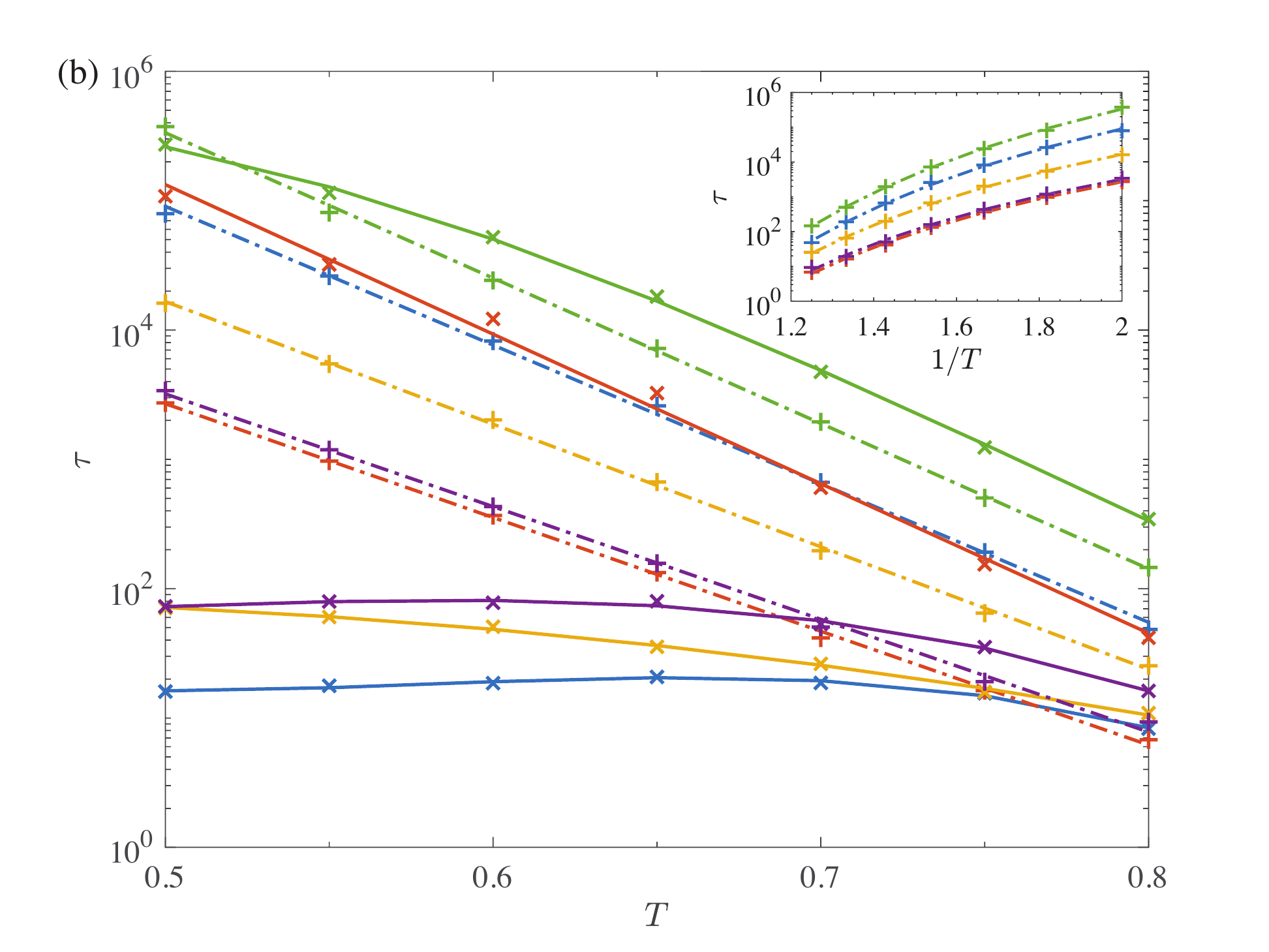}
\end{minipage}
}
\caption{Evolution of the structural relaxation time in the cluster fluid (a) with density at $T = 0.7$ and (b) with temperature at $\rho = 0.06$ using MMC (green), VMMC (blue), AVBMC (orange), ECMC (yellow) and VMMC/AVBMC combination (purple) algorithms. Both the single-particle $\tau_\mathrm{P}$ (from Eq.~\eqref{eq:cpair}, dot-dashed lines) and the collective $\tau_\mathrm{C}$ (from Eq.~\eqref{eq:ccluster}, solid lines) dynamics are considered. VMMC efficiently relaxes cluster positions ($\tau_\mathrm{C}$), but less so exchanges particles ($\tau_\mathrm{P}$). By contrast, AVBMC efficiently rearranges clusters  ($\tau_\mathrm{P}$), but is lackluster at enhancing cluster diffusion ($\tau_\mathrm{C}$). The inset presents the near Arrhenius scaling of the various schemes for an activation energy, $E_a\approx 9\varepsilon$.}\label{fig:dep}
\end{figure}

In cluster fluids, algorithmic efficiency is assessed from the decay time of the cluster reorganization correlation function $C_\mathrm{pair}(t)$ in Eq.~\eqref{eq:cpair}, and from the cluster position correlation function $C_\mathrm{clust}(t)$ in Eq.~\eqref{eq:ccluster}. In this and following cases, we denote these times as being either particle-based or collective, $\tau_\mathrm{P}$ and $\tau_\mathrm{C}$, respectively. 

Unsurprisingly, the various advanced MC methods introduced in Sec.~\ref{sec:MCmethods} are differently efficient in decorrelating these two processes (Fig.~\ref{fig:dep}). As expected, VMMC efficiently displaces clusters but inefficiently reorganizes them, while the opposite is true of AVBMC (see Fig.~\ref{fig:methods}). In this context, combining the two schemes is natural. The resulting efficiency is found to be largely independent of the precise ratio of VMMC and AVBMC attempted displacements, as long as both are used with comparable frequency. Using $20\%$ VMMC moves is found to be near optimal under a variety of conditions and is thus kept for the remainder of the study. 

The optimized combination algorithm exhibits a solid overall performance, exceeding that of standard MMC by two orders of magnitude. However,  the cluster reorganization efficiency of the combined algorithm is bounded by that of AVBMC, and the cluster diffusion by that of VMMC. In other words, the contribution of the two algorithms is additive rather than synergistic. Based on the density-dependent trends shown in Fig.~\ref{fig:dep}(a), even though cluster diffusion of VMMC is rapid in the tested region, the increase of corresponding $\tau_\mathrm{C}$ is nearly exponential and much faster than that of AVBMC. It therefore becomes rate-limiting as density increases, and if $\tau_\mathrm{C}$ for VMMC exceeds that for AVBMC, then the combination algorithm does not improve performance over that of AVBMC alone. 

The temperature dependence of $\tau_\mathrm{P}$  suggests an Arrhenius-like scaling with an activation energy $E_a \approx 9\varepsilon$, roughly corresponding to the net number of bonds broken by displacing a particle. By contrast, $\tau_\mathrm{C}$, which does not necessitate bond breaking (except for AVBMC), is largely temperature independent (Fig.~\ref{fig:dep}(b)). As a result, the difference between VMMC and AVBMC in $\tau_\mathrm{C}$ is larger at lower $T$, making the combined algorithm relatively more efficient. 

ECMC exhibits intermediate efficiency for both processes, and its $\tau_\mathrm{P}$ scales similarly to that of other methods. It is therefore never optimal at low $\rho$. Increasing density, however, diminishes the efficiency of ECMC more mildly than that of the other algorithms. Extrapolating the current trends suggests that ECMC should become optimal when this morphological regime to extend slightly further. It therefore cannot be excluded that ECMC could be optimal for sampling clusters in other SALR models. This should be particularly true for systems with large clusters, for which the relative surface contribution (useful for AVBMC moves) is diminished. 

\subsection{Percolated Fluid}
% figure: percolation dynamics
\begin{figure}
\centering
\includegraphics[width=8.5cm,trim={1cm 0.2cm 0 0.8cm},clip]{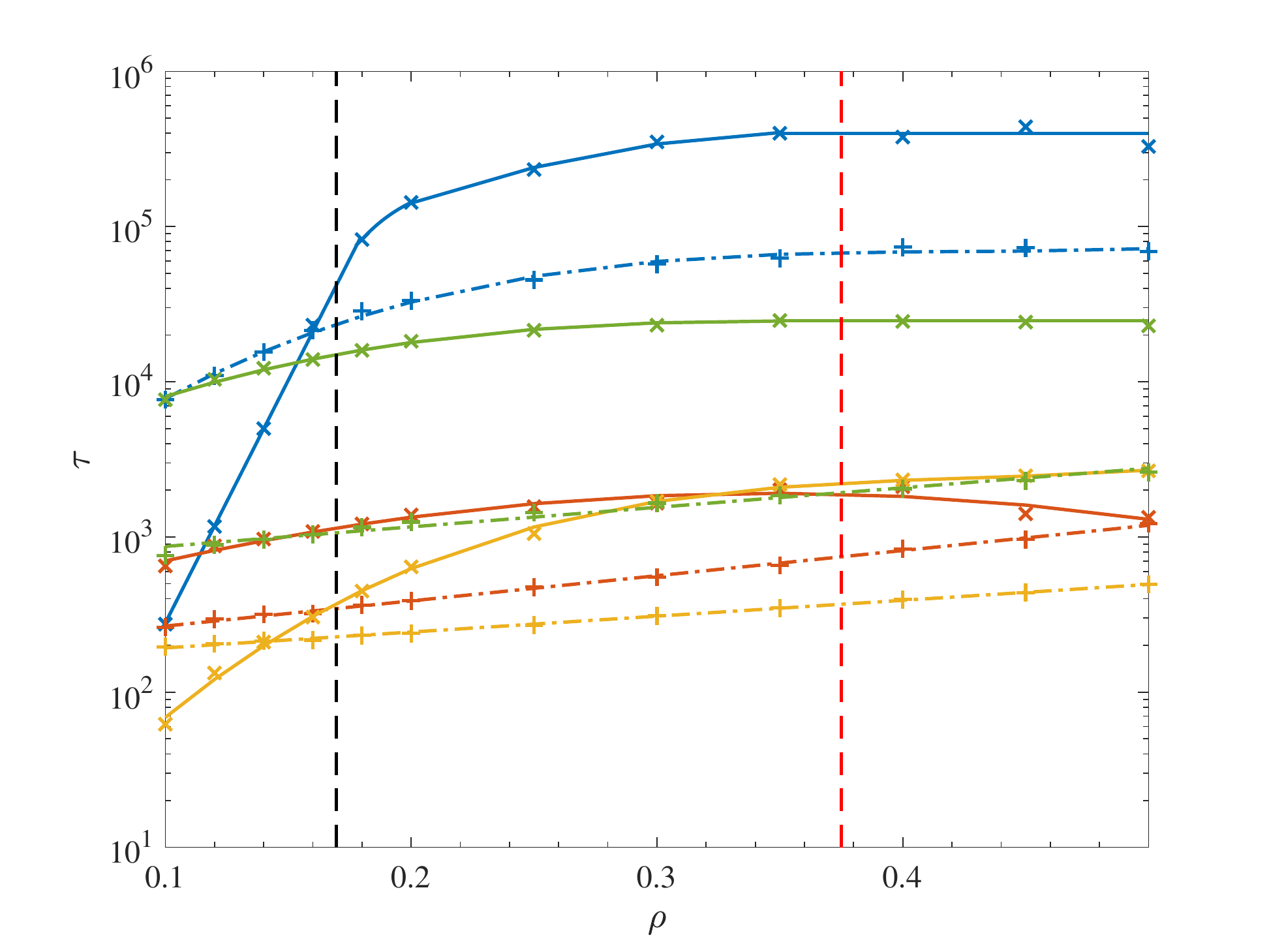}
\caption{Density evolution of structural relaxation time in the percolated fluid for $T = 0.7$ using MMC (green), VMMC (blue), AVBMC (orange) and ECMC (yellow). The single-particle  $\tau_\mathrm{P}$  %$C_\mathrm{bond}(t)$
(from Eq.~\eqref{eq:p_bond}, dot-dashed) and the collective $\tau_\mathrm{C}$ 
%$C_\mathrm{part}(t)$
(Eq.~\eqref{eq:p_particle}, solid lines) dynamics reveal that only VMMC is affected by percolation (black dashed line). The proximity or $\rho_\mathrm{ODT}$ (red dashed line) appears to maximize to collective slowdown, but the effect is largely insignificant already at this $T>T_\mathrm{ODT}$.}
\label{fig:percolation}
\end{figure}

Near and beyond the onset of physical percolation, the structural relaxation processes monitored by the local particle connections $C_\mathrm{bond}(t)$ (Eq.~\eqref{eq:p_bond}) and by the global network diffusion $C_\mathrm{ntwk}(t)$ (Eq.~\eqref{eq:p_particle}) (Fig.~\ref{fig:percolation}). As expected, VMMC and the combination algorithm described in Sec.~\ref{sec:clusdyn} becomes markedly less efficient in this regime. Cluster displacements are indeed largely unproductive as clusters themselves grow, elongate, and then become percolated. (VMMC is even slower than simple MMC for percolated systems.) Because the combination algorithm offers no improvement over AVBMC alone, it is not further considered. ECMC and AVBMC, which are essentially insensitive to percolation, nevertheless steadily become less efficient as $\rho$ increases. For ECMC, the effect is likely due to particle displacements being smaller in range as the fraction of void space steadily shrinks. At sufficiently high density, most particles hardly move before liftings, at which point the relaxation efficiency roughly saturates. AVBMC is more mildly affected by density because sufficient surface particles remain available in this regime.

Overall, ECMC is optimal around and beyond the percolation threshold. It accelerates dynamics by about one order of magnitude over MMC. As density increases, ECMC and AVBMC have comparable efficiencies. In a given SALR system, the fine balance between the two algorithms in this regime thus likely depends on the fraction of surface sites and on implementation considerations. 

\subsection{Void Cluster Fluid}
Near and beyond the void percolation threshold, particle networks are dense and liquid-like, and hence void clusters become the significant mesoscale features. The structural relaxation is then controlled by the liquid-state relaxation of particles $C_\mathrm{self}(t)$ (Eq.~\eqref{eq:v_self}) and that of void clusters $C_\mathrm{void}(t)$ (Eq.~\eqref{eq:v_void}), $\tau_P$ and $\tau_V$, respectively. Given the percolated fluid results, only ECMC and AVBMC are here compared to the standard MMC.

% figure: void dynamics
\begin{figure}
\centering
\includegraphics[width=8.5cm,trim={1cm 0.2cm 0 0.8cm},clip]{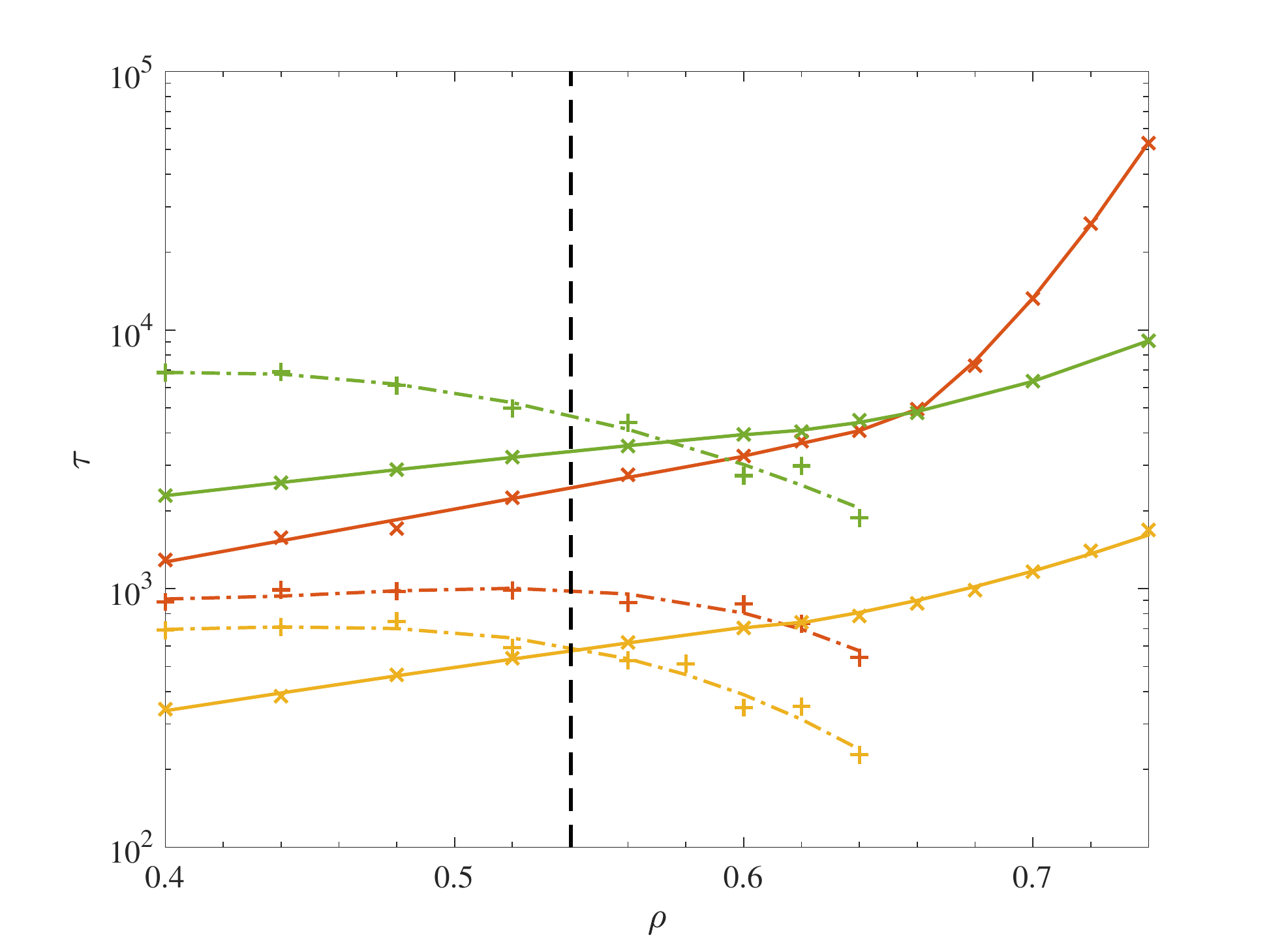}
\caption{Density evolution of the structural relaxation time in the void cluster fluid at $T = 0.7$, using the void $\tau_\mathrm{V}$ (from Eq.~\eqref{eq:v_void}, dot-dashed lines) and the particle $\tau_\mathrm{P}$ (from Eq.~\eqref{eq:v_self}, solid lines) dynamics for MMC (green), AVBMC (orange) and ECMC (yellow). The void percolation threshold (black dashed line) is given as reference. For $\rho > 0.68$, too few voids remain, hence $\tau_\mathrm{V}$ results are missing.}
\label{fig:voids}
\end{figure}

The void correlation times $\tau_\mathrm{V}$ of ECMC and AVBMC are fairly insensitive to crossing the void percolation threshold. Given that both algorithms rely on either single-particle or local updates, this independence to the global structure is not particularly surprising. As density further increases, however, a marked speed up can be observed. This follows from clusters diffusing faster as they shrink. Increasing density also reduces the number (volume) of void monomers. Beyond a given density ($\rho \sim 0.64$ in Fig.~\ref{fig:voids}), the very definition of void clusters becomes ambiguous, and from that point on only the particle behavior is considered. Note that although one might expect the void cluster decorrelation time to be larger than that of particles, given that void positions here only decorrelate when particles do, surface particles (in contact with voids) reorganize much faster than in the bulk, which explains the difference.

Particle decorrelation ($\tau_\mathrm{P}$) is thus the rate-limiting process for both algorithms at high densities. For AVBMC, the increase of $\tau_\mathrm{P}$ with densities is slow at first, but accelerates dramatically as the fraction of void space vanishes (for $\rho\sim0.64$ in Fig.~\ref{fig:voids}), because without voids surface swap moves become highly inefficient. By contrast, $\tau_\mathrm{P}$ for ECMC increases only slowly due its relative independence on the presence of voids. As a result, ECMC is the most efficient algorithm in the void cluster regime, accelerating the relaxation process by one order of magnitude over MMC, over a wide density range. 

\section{Conclusion}
\label{sec:conclusion}
Our consideration of a schematic SALR model has allowed us to use numerical methods to resolve the various structural features of its disordered microphases. Rich morphologies, such as  cluster and percolated fluids are observed at temperatures intermediate between ordered microphases (at low temperatures) and truly homogeneous fluids (at high temperatures). Like their ordered counterparts, disordered microphases display a rough symmetry between particles and voids, thus echoing lattice models~\cite{tarzia2021}. We find that their thermal features, however, do not similarly match, likely because of the abundance of particle-based degrees of freedom at high densities. 

Evaluating the efficiency of various advanced MC algorithms--either isolated or in combination--for the various morphological regimes provides insight into the interplay between these algorithms and structure. In particular, for the SALR model considered, we find that a combination algorithm of VMMC and AVBMC is most efficient for cluster fluids, which improves the sampling efficiency over standard local sampling algorithms by at least two orders of magnitude. 
Both AVBMC and ECMC are roughly equivalent in the percolated fluid, and ECMC is most efficient in the void cluster regime. Overall, the optimal algorithm offers about one order of magnitude speed up over MMC. 
These results should thus both guide future numerical studies of microphase formers as well as establish benchmarks for future algorithm development efforts for these challenging systems. 

Various numerical challenges remain in characterizing disordered microphases. For instance, considering systems with a shorter attraction range would be more representative of depletion attraction in colloidal suspensions, but face serious sampling hurdles. An efficient numerical determination of the ODT, at which point structures become more compact and relax more slowly, also largely remains an open problem. 

\begin{acknowledgments}
We thank  Yi Hu and Marco Tarzia for stimulating discussions as well as for sharing results ahead of publication. This work was supported by a the National Science Foundation Grant No.~DMR-1749374. Data relevant to this work have been archived and can be accessed at the Duke Digital Repository.\cite{mfdata}
\end{acknowledgments}

%\vspace{2cm}

\appendix
\section{Virial coefficients}
\label{sec:appendix}
\newmuskip\pFqmuskip
\newcommand*\pFq[6][8]{
  \begingroup 
  \pFqmuskip=#1mu\relax
  \mathchardef\normalcomma=\mathcode`,
  % make the comma math active
  \mathcode`\,=\string"8000
  % and define it to be \pFqcomma
  \begingroup\lccode`\~=`\,
  \lowercase{\endgroup\let~}\pFqcomma
  % typeset the formula
  {}_{#2}F_{#3}{\left[\genfrac..{0pt}{}{#4}{#5};#6\right]}
  \endgroup
}
\newcommand{\pFqcomma}{{\normalcomma}\mskip\pFqmuskip}

In the context of Eq.~\eqref{eq:h_rho}, and because of the absence of thermodynamic singularities in the cluster fluid regime, it is instructive to express the equilibrium pressure as a virial series
\begin{equation}
    \beta P = \rho + \sum_{k=2}^\infty B_n(T)\rho^n
\end{equation}
with coefficients $B_n(T)$. In this appendix, we obtain analytical expressions for $B_2(T)$, $B_3(T)$ and $B_4(T)$ for the generic SALR model specified by Eq.~\eqref{eq:hamiltonian} and in arbitrary spatial dimension $d$. Because few computations of virial coefficients exist for SALR models (and none for $B_4(T)$), these results are particularly informative for understanding the low-density clustering behavior. 

Here we follow the derivation approach outlined in Refs.~\onlinecite{mccoy2010advanced,santos2016concise}. For the model in Eq.~\eqref{eq:hamiltonian}, the Mayer function is
\begin{align}
     f(r) &=e^{-\beta u(r)}-1\\
     &= \left\{ \begin{array}{llll}
              -1 & \mbox{$r < \sigma $};\\
             f_0 & \mbox{$\sigma < r < \lambda_0\sigma$};\\
             f_1 & \mbox{$\lambda_0\sigma < r < \lambda_1\sigma$};\\
             0 & \mbox{$\lambda_1\sigma < r$}.
             \end{array} \right. 
\end{align}
where $f_0 = e^{\beta\varepsilon}-1$ and $f_1=e^{-\beta\kappa\varepsilon}-1$, and its Fourier transform is
\begin{align}
\label{eqn:fourier}
    F_d(k) &= \int e^{-i\mathbf{k}\cdot\mathbf{r}}f(r)d\mathbf{r}\nonumber\\
    &= \left(\frac{2\pi\sigma}{k}\right)^{d/2}[-(1+f_0)J_{d/2}(k\sigma)\\
    &+(f_0-f_1)\lambda_0^{d/2}J_{d/2}(k\lambda_0\sigma)
    +f_1\lambda_1^{d/2}J_{d/2}(k\lambda_1\sigma)].\nonumber
\end{align}

For analytical convenience, we recall a few relations. \begin{itemize}
    \item For a spherically symmetric function $H(r)$
\begin{equation}
\label{eqn:int1}
    \int H(r)d\mathbf{r} = C_d\int_0^\infty r^{d-1}H(r)dr
\end{equation}
where $C_d = \frac{2\pi^{d/2}}{\Gamma(d/2)}$. 
\item For a function $H$ of distance $r$ and of a single polar angle $\theta$ (such as $H(r,\theta) = e^{-ik\cdot r\cos{\theta}}f(r)$ in Eq.~\eqref{eqn:fourier})
\begin{align}
\label{eqn:int2}
\int H(r,\theta)d\mathbf{r}= C_{d-1}\int_0^\infty\!\!\!\! dr\,r^{d-1}\!\!\!\int_0^\pi\!\!\! d\theta\,(\sin{\theta})^{d-2} H(r,\theta).
\end{align}
\item The Bessel function in Eq.~\eqref{eqn:fourier} can be expressed as \cite{erdelyi1972m}
\begin{equation}
    J_\nu(x) = \frac{(x/2)^\nu}{\sqrt{\pi}\Gamma(\nu+1/2)}
\int_0^\pi d\theta (\sin{\theta})^{2\nu}e^{ix\cos{\theta}}
\end{equation}
for $\mathrm{Re}(\nu)>-\frac{1}{2}$, and obeys the relation
\begin{equation}
    \frac{d}{dx}[x^{\nu}J_\nu (x)] = x^\nu J_{\nu-1}(x).
\end{equation}
\end{itemize}

\subsection{Second Virial Coefficient}
Using Eq.~\eqref{eqn:int1}, the standard expression for $B_2(T)$ is obtained
\begin{align}
    B_2(T) &= -\frac{1}{2}\int f(r)d\mathbf{r}\nonumber\\
    &= \frac{C_d\sigma^d}{2d}[1+f_0-(f_0-f_1)\lambda_0^d-f_1\lambda_1^d].
\end{align}

% third
\subsection{Third Virial Coefficient}
Expressing $B_3(T)$ in Fourier space and using Eq.~\eqref{eqn:fourier} immediately gives
\begin{widetext}
\begin{align*}
\label{eqn:b3_fourier}
    B_3(T) &= -\frac{1}{3}\int \! \! \! \int f(r_1)f(r_2)
    f(|\mathbf{r_2}-\mathbf{r_1}|)\,d\mathbf{r_1}\,d\mathbf{r_2}= -\frac{1}{3 (2\pi)^d}\int [F_d(k)]^3\,d\mathbf{k}\\
    &= -\frac{C_d}{3}(2\pi)^{d/2}\sigma^{2d}\lbrace[-(1+f_0)^3+(f_0-f_1)^3\lambda_0^{2d}+f_1^3\lambda_1^{2d}]I^{(3)}_1+[-3(1+f_0)(f_0-f_1)^2\lambda_0^d]I^{(3)}_2+[-3(1+f_0)f_1^2\lambda_1^d]I^{(3)}_3\\
    &+[3(1+f_0)^2(f_0-f_1)\lambda_0^{d/2}]I^{(3)}_4+[3(1+f_0)^2f_1\lambda_1^{d/2}]I^{(3)}_5+[3(f_0-f_1)^2f_1\lambda_0^d\lambda_1^{d/2}]I^{(3)}_6+[3(f_0-f_1)f_1^2\lambda_0^{d/2}\lambda_1^d]I^{(3)}_7\\
    & +[-3(1+f_0)(f_0-f_1)f_1\lambda_0^{d/2}\lambda_1^{d/2}]I^{(3)}_8\rbrace,
\end{align*}
\end{widetext}
where various integrals of Bessel functions, $I_i$, have been implicitly defined, such as
\begin{equation}
    I^{(3)}_1 = \int_0^\infty x^{-d/2-1}[J_{d/2}(x)]^3\,dx.
\end{equation}
Reference~[\onlinecite{luban1982third}] provides expressions for these integrals. In particular,
\begin{align}
    &\int_0^\infty x^{-(1+\nu)}[J_\nu(x)]^3\,dx \\
    &= \frac{3\cdot 2^{-(\nu+2)}}{\sqrt{\pi}\nu\Gamma(\nu+1/2)}
    \cdot \left\{B\left(\frac{1}{2},\frac{1}{2}+\nu\right)-\pFq{2}{1}{\frac{1}{2},\frac{1}{2}-\nu}{
    \frac{3}{2}}{\frac{1}{4}}\right\} \nonumber
\end{align}
using the beta function, $B(x,y) = \Gamma(x)\Gamma(y)/\Gamma(x+y)$, and the hypergeometric functions $\pFq{p}{q}{a_1,\cdots,a_p}{b_1,\cdots,b_q}{c}$, which can be calculated from their power-series representations. High-accuracy values can thus be straightforwardly obtained for the following expressions
\begin{align}
        I^{(3)}_1 &= \frac{3 \times 2^{-d/2-1}}{d\sqrt{\pi}\Gamma(\frac{d+1}{2})} \nonumber\\
    &\times \left\{ B\left(\frac{1}{2},\frac{d+1}{2}\right)-\pFq{2}{1}{\frac{1}{2},\frac{1-d}{2}}{\frac{3}{2}}{\frac{1}{4}}\right\}.
\end{align}\\
From the special integrals given in Refs.~[\onlinecite{hussein1991virial,prudnikov2018integrals}], the other integrals can also be obtained:
\begin{widetext}
\begin{equation}
    I^{(3)}_2 = -\frac{2^{-d/2-1}}{\lambda_0\sqrt{\pi}\Gamma(\frac{3+d}{2})}
    \cdot \pFq{3}{2}{\frac{1-d}{2},\frac{1}{2},\frac{1+d}{2}}{\frac{3}{2},\frac{3+d}{2}}{\frac{1}{4\lambda_0^2}} + \frac{2^{-d/2+1}}{d^2\Gamma(d/2)} ;
\end{equation}
\begin{equation}
    I^{(3)}_3 = -\frac{2^{-d/2-1}}{\lambda_1\sqrt{\pi}\Gamma(\frac{3+d}{2})}
    \cdot \pFq{3}{2}{\frac{1-d}{2},\frac{1}{2},\frac{1+d}{2}}{\frac{3}{2},\frac{3+d}{2}}{\frac{1}{4\lambda_1^2}} + \frac{2^{-d/2+1}}{d^2\Gamma(d/2)} ;
\end{equation}
\begin{equation}
     I^{(3)}_4 = \begin{cases}
    -\frac{2^{-d/2-1}\lambda_0^{d/2+1}}{\sqrt{\pi}\Gamma(\frac{3+d}{2})}
    \cdot \pFq{3}{2}{\frac{1-d}{2},\frac{1}{2},\frac{1+d}{2}}{\frac{3}{2},\frac{3+d}{2}}{\frac{\lambda_0^2}{4}} + \frac{2^{-d/2+1}\lambda_0^{d/2}}{d^2\Gamma(d/2)},
    & \mbox{$\lambda_0 < 2$}; \\
    \frac{2^{-d/2+1}\lambda_0^{-d/2}}{d^2\Gamma(d/2)}, 
    & \mbox{$\lambda_0 > 2$}; \\
    \end{cases}
\end{equation}
\begin{equation}
     I^{(3)}_5 = \begin{cases}
    -\frac{2^{-d/2-1}\lambda_1^{d/2+1}}{\sqrt{\pi}\Gamma(\frac{3+d}{2})}
    \cdot \pFq{3}{2}{\frac{1-d}{2},\frac{1}{2},\frac{1+d}{2}}{\frac{3}{2},\frac{3+d}{2}}{\frac{\lambda_1^2}{4}} + \frac{2^{-d/2+1}\lambda_1^{d/2}}{d^2\Gamma(d/2)},
    & \mbox{$\lambda_1 < 2$};\\
    \frac{2^{-d/2+1}\lambda_1^{-d/2}}{d^2\Gamma(d/2)}, 
    & \mbox{$\lambda_1 > 2$};\\
    \end{cases}
\end{equation}
\begin{equation}
     I^{(3)}_6 = \begin{cases}
    -\frac{2^{-d/2-1}\lambda_1^{d/2+1}}{\lambda_0\sqrt{\pi}\Gamma(\frac{3+d}{2})}
    \cdot \pFq{3}{2}{\frac{1-d}{2},\frac{1}{2},\frac{1+d}{2}}{\frac{3}{2},\frac{3+d}{2}}{\frac{\lambda_1^2}{4\lambda_0^2}} + \frac{2^{-d/2+1}\lambda_1^{d/2}}{d^2\Gamma(d/2)},
    & \mbox{$\lambda_1 < 2\lambda_0$};\\
    \frac{2^{-d/2+1}\lambda_0^d\lambda_1^{-d/2}}{d^2\Gamma(d/2)},
    & \mbox{$\lambda_1 > 2\lambda_0$};\\
    \end{cases}
\end{equation}
\begin{equation}
    I^{(3)}_7 = -\frac{2^{-d/2-1}\lambda_0^{d/2+1}}{\lambda_1\sqrt{\pi}\Gamma(\frac{3+d}{2})}
    \cdot \pFq{3}{2}{\frac{1-d}{2},\frac{1}{2},\frac{1+d}{2}}{\frac{3}{2},\frac{3+d}{2}}{\frac{\lambda_0^2}{4\lambda_1^2}} + \frac{2^{-d/2+1}\lambda_0^{d/2}}{d^2\Gamma(d/2)} ;
\end{equation}
\begin{equation}
    I^{(3)}_8 = \left(\frac{\lambda_0}{\lambda_1}\right)^{d/2} \frac{2^{-d/2+1}}{d^2\Gamma(d/2)} \quad \mbox{for $\lambda_1 > \lambda_0 + 1$}.
\end{equation}
\end{widetext}

\subsection{Fourth Virial Coefficient}
The fourth virial coefficient has three Mayer cluster contributions 
\begin{equation}
    B_4(T) = B_4(\Square)+B_4(\boxbslash)+B_4(\XBox).
\end{equation}
These terms, however, no longer have generic closed-form expressions for arbitrary $d$.
The integrals can nevertheless be analytically expressed for specific odd dimensions. The derivation below follows the strategy used in Ref.~[\onlinecite{hussein1991virial}].

%B4_1
\emph{First Term:} 
In Fourier representation we have 
\begin{widetext}
\begin{equation}
\begin{split}
B_4(\Square) &= -\frac{3}{8}\int \! \! \! \int \! \! \! \int
f(r_1)f(r_2)f(|\mathbf{r_3}-\mathbf{r_2}|)f(|\mathbf{r_1}-\mathbf{r_3}|)
\,d\mathbf{r_1}\,d\mathbf{r_2}\,d\mathbf{r_3}= -\frac{3}{8}\cdot (2\pi)^{-d}\int [F_d(k)]^4\,d\mathbf{k}\\
     &= -\frac{3}{8}C_d(2\pi)^d\sigma^{3d}\lbrace[(1+f_0)^4+(f_0-f_1)^4\lambda_0^{3d}+f_1^4\lambda_1^{3d}]I_1^{(\Square)} +[-4(1+f_0)^3(f_0-f_1)\lambda_0^{d/2}]I_2^{(\Square)}\\
    &+[-4(1+f_0)^3f_1\lambda_1^{d/2}]I_3^{(\Square)}
    +[-4(1+f_0)(f_0-f_1)^3\lambda_0^{3d/2}]I_4^{(\Square)}
    +[-4(1+f_0)f_1^3\lambda_1^{3d/2}]I_5^{(\Square)}\\
    &+[4(f_0-f_1)^3f_1\lambda_0^{3d/2}\lambda_1^{d/2}]I_6^{(\Square)}
    +[4(f_0-f_1)f_1^3\lambda_0^{d/2}\lambda_1^{3d/2}]I_7^{(\Square)}
    +[6(1+f_0)^2(f_0-f_1)^2\lambda_0^d]I_8^{(\Square)}\\
    &+[6(1+f_0)^2f_1^2\lambda_1^d]I_9^{(\Square)}
    +[6(f_0-f_1)^2f_1^2\lambda_0^d\lambda_1^d]I_{10}^{(\Square)}
    +[12(1+f_0)^2(f_0-f_1)f_1\lambda_0^{d/2}\lambda_1^{d/2}]I_{11}^{(\Square)}\\
    &+[-12(1+f_0)(f_0-f_1)^2f_1\lambda_0^d\lambda_1^{d/2}]I_{12}^{(\Square)}
    +[-12(1+f_0)(f_0-f_1)f_1^2\lambda_0^{d/2}\lambda_1^d]I_{13}^{(\Square)}
    \rbrace ,
\end{split}
\end{equation}
\end{widetext}

where the integrals
\begin{equation}
I_1^{(\Square)} = \int_0^\infty x^{-d-1}[J_{d/2}(x)]^4\,dx
\end{equation}
\begin{equation}
I_2^{(\Square)} = \int_0^\infty x^{-d-1}[J_{d/2}(x)]^3J_{d/2}(\lambda_0x)\,dx
\end{equation}
\begin{equation}
I_3^{(\Square)} = \int_0^\infty x^{-d-1}[J_{d/2}(x)]^3J_{d/2}(\lambda_1x)\,dx
\end{equation}
\begin{equation}
I_4^{(\Square)} = \int_0^\infty x^{-d-1}J_{d/2}(x)[J_{d/2}(\lambda_0x)]^3\,dx
\end{equation}
\begin{equation}
I_5^{(\Square)} = \int_0^\infty x^{-d-1}J_{d/2}(x)[J_{d/2}(\lambda_1x)]^3\,dx
\end{equation}
\begin{equation}
I_6^{(\Square)} = \int_0^\infty x^{-d-1}[J_{d/2}(\lambda_0x)]^3J_{d/2}(\lambda_1x)\,dx
\end{equation}
\begin{equation}
I_7^{(\Square)} = \int_0^\infty x^{-d-1}J_{d/2}(\lambda_0x)[J_{d/2}(\lambda_1x)]^3\,dx
\end{equation}
\begin{equation}
I_8^{(\Square)} = \int_0^\infty x^{-d-1}[J_{d/2}(x)]^2[J_{d/2}(\lambda_0x)]^2\,dx
\end{equation}
\begin{equation}
I_9^{(\Square)} = \int_0^\infty x^{-d-1}[J_{d/2}(x)]^2[J_{d/2}(\lambda_1x)]^2\,dx
\end{equation}
\begin{equation}
I_{10}^{(\Square)} = \int_0^\infty x^{-d-1}[J_{d/2}(\lambda_0x)]^2[J_{d/2}(\lambda_1x)]^2\,dx
\end{equation}
\begin{equation}
I_{11}^{(\Square)} = \int_0^\infty x^{-d-1}[J_{d/2}(x)]^2J_{d/2}(\lambda_0x)J_{d/2}(\lambda_1x)\,dx
\end{equation}
\begin{equation}
I_{12}^{(\Square)} = \int_0^\infty x^{-d-1}J_{d/2}(x)[J_{d/2}(\lambda_0x)]^2J_{d/2}(\lambda_1x)\,dx
\end{equation}
\begin{equation}
I_{13}^{(\Square)} = \int_0^\infty x^{-d-1}J_{d/2}(x)J_{d/2}(\lambda_0x)[J_{d/2}(\lambda_1x)]^2\,dx
\end{equation}\\
For a given $d$, these expression can be analytically evaluated using symbolic mathematical tools, such as Mathematica.

%B4_2
\emph{Second Term:} Again using the Fourier representation, we have
\begin{widetext}
\begin{equation}
\begin{split}
B_4(\boxbslash) &= -\frac{3}{4}\int \! \! \! \int \! \! \! \int
f(r_1)f(r_2)f(r_3)f(|\mathbf{r_1}-\mathbf{r_2}|)f(|\mathbf{r_2}-\mathbf{r_3}|)\,d\mathbf{r_1}\,d\mathbf{r_2}\,d\mathbf{r_3}\\
&= -\frac{3}{4}\cdot (2\pi)^{-2d}\int [F_d(k)F_d(k')]^2f(r)e^{-i(\mathbf{k}+\mathbf{k'})\cdot\mathbf{r}}\,d\mathbf{k}\,d\mathbf{k'}\,d\mathbf{r}\\
&= -\frac{3}{4}\cdot (2\pi)^{-2d} 
\left\{ -\int_0^\sigma\left[\int_0^\infty[F_d(k)]^2e^{-i\mathbf{k}\cdot\mathbf{r}}\,d\mathbf{k}\right]^2\,d\mathbf{r}
+ f_0\int_\sigma^{\lambda_0\sigma}\left[\int_0^\infty[F_d(k)]^2e^{-i\mathbf{k}\cdot\mathbf{r}}\,d\mathbf{k}\right]^2\,d\mathbf{r}\right. \\
&+ \left. f_1\cdot\int_{\lambda_0\sigma}^{\lambda_1\sigma}\left[\int_0^\infty[F_d(k)]^2e^{-i\mathbf{k}\cdot\mathbf{r}}\,d\mathbf{k}\right]^2\,d\mathbf{r} \right\} \\
\end{split}
\end{equation}
The three terms containing multiple integrals should be handled similarly. Taking the first term as an example, we get
\begin{equation}
\begin{split}
    I_1^{(\boxbslash)} &= -\int_0^\sigma\left[\int_0^\infty[F_d(k)]^2e^{-i\mathbf{k}\cdot\mathbf{r}}\,d\mathbf{k}\right]^2\,d\mathbf{r}\\
    &= -\int_0^\sigma\left[C_{d-1}\int_0^\infty\int_0^\pi [F_d(k)]^2\cdot e^{-ikr\cos{\theta}}k^{d-1}(\sin{\theta})^{d-2}\,dk\,d\theta\right]^2\cdot C_d\cdot r^{d-1}\,dr\\
\end{split}
\end{equation}

\noindent The integral over ($dk\,d\theta$) gives
\begin{equation}
\begin{split}
    K_1 &= C_{d-1}\int_0^\infty[F_d(k)]^2k^{d-1}\,dk\int_0^\pi e^{-ikr\cos{\theta}}(\sin{\theta})^{d-2}\,d\theta\\
    &= (2\pi)^{3d/2}\sigma^d y^{-d/2+1}\int_0^\infty J_{d/2-1}x^{-d/2}\lbrace (1+f_0)^2[J_{d/2}(x)]^2\\
    &+(f0-f1)^2\lambda_0^d[J_{d/2}(\lambda_0x)]^2 
    + f_1^2\lambda_1^d[J_{d/2}(\lambda_1x)]^2 \\
    &- 2(1+f_0)(f_0-f_1)\lambda_0^{d/2}J_{d/2}(x)J_{d/2}(\lambda_0x) - 2(1+f_0)f_1\lambda_1^{d/2}J_{d/2}(x)J_{d/2}(\lambda_1x)\\
    &+ 2(f_0-f_1)f_1\lambda_0^{d/2}\lambda_1^{d/2}J_{d/2}(\lambda_0x)J_{d/2}(\lambda_1x)\rbrace\,dx
\end{split}
\end{equation}
\end{widetext}

\noindent where $x=k\sigma$, $r=\sigma y$ and $kr=xy$. Conveniently, the integrals involving $\mathbf{k}$, i.e., $k$ and $\theta$, in the three terms of $B_4(\boxbslash)$ share a same expression. Then the expressions can be simplified with the outer integration over $y$. 
For the first term, $y\in[0,1]$; for the second term, $y\in[1,\lambda_0]$; for the third term, $y\in[\lambda_0,\lambda_1]$. 

\begin{equation}
I_1^{(\boxbslash)} = - C_d \sigma^d\int_0^1 K_1^2\cdot y^{d-1}\,dy
\end{equation}
\begin{equation}
I_2^{(\boxbslash)} = f_0 \cdot C_d \sigma^d\int_1^{\lambda_0} K_2^2\cdot y^{d-1}\,dy\
\end{equation}
\begin{equation}
I_3^{(\boxbslash)} = f_1 \cdot C_d \sigma^d\int_{\lambda_0}^{\lambda_1} K_3^2\cdot y^{d-1}\,dy
\end{equation}
\noindent The integrals of Bessel functions for any definite dimension can be calculated using Mathematica.\\
\\

%B4_3
\emph{Third Term:} This term is the most complicated contribution to $B_4$, because it involves all types of pair interactions in a group of four particles, from the perspective of diagrammatic method.
\begin{widetext}
\begin{equation}\label{eq:b4_3}
\begin{split}
    B_4(\XBox) &= -\frac{1}{8}\int \! \! \! \int \! \! \! \int
f(r_1)f(r_2)f(r_3)f(|\mathbf{r_1}-\mathbf{r_2}|)f(|\mathbf{r_2}-\mathbf{r_3}|)f(|\mathbf{r_3}-\mathbf{r_1}|)\,d\mathbf{r_1}\,d\mathbf{r_2}\,d\mathbf{r_3}\\
&= -\frac{1}{8}(2\pi)^{-3d}\int \! \! \! \int \! \! \! \int F_d(k)F_d(k')F_d(k'')F_d(|\mathbf{k''}-\mathbf{k}|)F_d(|\mathbf{k}-\mathbf{k'}|)F_d(|\mathbf{k'}-\mathbf{k''}|)\,d\mathbf{k}\,d\mathbf{k'}\,d\mathbf{k''}\\
&= -\frac{1}{8}\cdot \sigma^{3d}\sum_{a_1\cdots a_6}[-(1+f_0)]^{n_1}[(f_0-f_1)\lambda_0^d]^{n_2}[f_1\lambda_1^d]^{n_3}
\times I_{d/2}(a_1,a_2,a_3,a_4,a_5,a_6)
\end{split}
\end{equation}

\noindent where $x = \sigma k$, $a_1\cdots a_6\in\lbrace 1,\lambda_0,\lambda_1\rbrace$, and $n_1$, $n_2$ and $n_3$  represents the number of $a_i$ equal to 1, $\lambda_0$ or $\lambda_1$, respectively.\\
\begin{equation}\label{eq:b4_3_I}
\begin{split}
&I_{d/2}(a_1,a_2,a_3,a_4,a_5,a_6) \\
&= \int \! \! \! \int \! \! \! \int h_{d/2}(a_1x)h_{d/2}(a_2x')h_{d/2}(a_3x'')h_{d/2}(a_4|\mathbf{x''}-\mathbf{x}|)h_{d/2}(a_5|\mathbf{x}-\mathbf{x'}|)h_{d/2}(a_6|\mathbf{x'}-\mathbf{x''}|)\,d\mathbf{x}\,d\mathbf{x'}\,d\mathbf{x''} 
\end{split}
\end{equation}
\noindent where $h_{d/2}(t)=\frac{J_{d/2}(t)}{t^{d/2}}$.\\

\noindent From Ref.~\onlinecite{bateman1953higher}[Vol.~2, p.~101, Eq.~(30)], the addition theorem of Bessel function gives 
\begin{equation}
\begin{split}
w^{-\mu}J_\mu(w) = (\frac{1}{2}zZ)^{-\mu}\Gamma(\mu)\sum_{n=0}^\infty(\mu+n)C_n^\mu(\cos{\nu})J_{\mu+n}(z)J_{\mu+n}(Z) \qquad\mbox{for $\mu\not= 0, -1, \cdots$},
\end{split}
\end{equation}
where
\begin{equation}
w = (z^2+Z^2-2zZ\cos{\nu})^{1/2}
\end{equation}
and using Gegenbauer polynomials defined as 
\begin{equation}
C_n^{\mu}(z) = 2^{\mu-1/2}\frac{\Gamma(n+2\mu)\Gamma(\mu+1/2)}{\Gamma(2\mu)\Gamma(n+1)}(z^2-1)^{1/4-1/2\mu}\cdot P_{n+\mu-1/2}^{1/2-\mu}(z).
\end{equation}
As a result,
\begin{equation}
\begin{split}
    h(a|\mathbf{x}-\mathbf{x'}|) &= \frac{J_{d/2}(a|\mathbf{x}-\mathbf{x'}|)}{(a|\mathbf{x}-\mathbf{x'}|)^{d/2}}
    = 2^{d/2}a^{-d}(xx')^{-d/2}\Gamma(d/2)\sum_{n=0}^\infty(d/2+n)C_n^{d/2}(\cos{\nu})J_{d/2+n}(ax)J_{d/2+n}(ax')
\end{split}
\end{equation}
and
\begin{equation}\label{eq:b4_3_I2}
\begin{split}
    & I_{d/2}(a_1,a_2,a_3,a_4,a_5,a_6)\\
    &= [2^{d/2}\Gamma(d/2)]^3(a_1a_2a_3)^{-1/2}(a_4a_5a_6)^{-1}\times \sum_{l=0}^\infty\sum_{m=0}^\infty\sum_{n=0}^\infty(d/2+l)(d/2+m)(d/2+n)\\
    &\times C_l^{d/2}(\cos{\nu})C_m^{d/2}(\cos{\nu'})C_n^{d/2}(\cos{\nu''})
    \times\int \! \! \! \int \! \! \! \int\,d\mathbf{x}\,d\mathbf{x'}\,d\mathbf{x''}\cdot
    J_{d/2}(a_1x)J_{d/2+l}(a_4x)J_{d/2+m}(a_5x)x^{-3d/2}\\
    &\times J_{d/2}(a_2x')J_{d/2+m}(a_5x')J_{d/2+n}(a_6x')x'^{-3d/2}
    \times J_{d/2}(a_3x'')J_{d/2+l}(a_4x'')J_{d/2+n}(a_6x'')x''^{-3d/2}\\
\end{split}
\end{equation}

In principle, $B_4(\XBox)$ for arbitrary odd dimension can be calculated from Eq.~\eqref{eq:b4_3} and Eq.~\eqref{eq:b4_3_I2}. Here, we only provide simplified expressions for $d=1$ and $d=3$.

\emph{In $d=1$:} Because $\cos{\nu}=1$, then $C_n^{1/2}(\cos{\nu})=1$ for any non-negative integer $n$, and thus
\begin{equation}
\begin{split}
    & I_{1/2}(a_1,a_2,a_3,a_4,a_5,a_6)\\
    &= (2\pi)^{3/2}(a_1a_2a_3)^{-1/2}(a_4a_5a_6)^{-1}\times \sum_{l=0}^\infty\sum_{m=0}^\infty\sum_{n=0}^\infty(1/2+l)(1/2+m)(1/2+n)\\
    &\times\int_0^\infty J_{1/2}(a_1x)J_{1/2+l}(a_4x)J_{1/2+m}(a_5x)x^{-3/2}\,dx\\
    &\times\int_0^\infty J_{1/2}(a_2x')J_{1/2+m}(a_5x')J_{1/2+n}(a_6x')x'^{-3/2}\,dx'\\
    &\times\int_0^\infty J_{1/2}(a_3x'')J_{1/2+l}(a_4x'')J_{1/2+n}(a_6x'')x''^{-3/2}\,dx''
\end{split}
\end{equation}\\

\emph{In $d=3$:} Based on Ref.~\onlinecite{katsura1959fourth}, we have
\begin{equation}
\begin{split}
    C_n^{3/2}(\cos{\nu})
    &= \frac{d P_{n+1}(\cos{\nu})}{d\cos{\nu}}\\
    &= (2n+1)P_n(\cos{\nu})+[2(n-2)+1]P_{n-2}(\cos{\nu})+[2(n-4)+1]P_{n-4}(\cos{\nu})+\cdots\\
    &= \sum_{u=\mbox{\scriptsize{even or odd}}, u\geq0}^{n} (2u+1)P_u(\cos{\nu})
\end{split}
\end{equation}
using the relation $(2n+1)P_n(x)=\frac{d}{dx}(P_{n+1}(x)-P_{n-1}(x)$. Moreover, 
\begin{equation}
    \,d\textbf{x} = x^2\,dx\,d\Omega=x^2\sin{\sigma}\,dx\,d\sigma\,d\psi
\end{equation}
where $\Omega$ is the solid angle. For $\cos\nu=\mathbf{x}\cdot\mathbf{x'}/(|x||x'|)$, we can also write
\begin{equation}
    \cos{\nu}=\cos{\sigma}\cos{\sigma'}+\sin{\sigma}\sin{\sigma'}\cos{(\psi-\psi')}
\end{equation}
We first deal with the angular part of the integral,
\begin{equation}
\begin{split}
    &\int C_l^{3/2}(\cos{\nu})C_m^{3/2}(\cos{\nu'})C_n^{3/2}(\cos{\nu''})\,d\Omega\,d\Omega'\,d\Omega''\\
    &= \int\sum_{u,v,w=}^l\sum_{\mbox{\scriptsize{even or odd,}}}^m\sum_{u,v,w\geq0}^n(2u+1)(2v+1)(2w+1)P_u(\cos{\nu})P_v(\cos{\nu'})P_w(\cos{\nu''})\,d\Omega\,d\Omega'\,d\Omega''\\
    &= \int\sum_u^{\min(l,m)}\sum_w^n(2u+1)^2(2w+1)\frac{4\pi}{2u+1}P_u(\cos{\nu''})P_w(\cos{\nu''})\,d\Omega\,d\Omega''\\
    &= \sum_u^{\min(l,m,n)}(2u+1)(4\pi)^2\int\,d\Omega''\\
    &= (4\pi)^3\sum_{u=\mbox{\scriptsize{even or odd}}, u\geq0}^{\min(l,m,n)}(2u+1)\\
    &= \frac{1}{2}(M+1)(M+2)(4\pi)^3
\end{split}
\end{equation}
\noindent where $M=\min(l,m,n)$, and
\begin{equation}
    \int P_u(\cos{\nu})P_v(\cos{\nu'})\,d\Omega' =
    \left\{
    \begin{array}{ll}
    4\pi P_u(\cos{\nu''})/(2u+1) & \mbox{if $u=v$};\\
    0 & \mbox{otherwise}.
    \end{array} \right. 
\end{equation}
\noindent As a result,
\begin{equation}
\begin{split}
    & I_{3/2}((a_1,a_2,a_3,a_4,a_5,a_6)\\
    &= 4\cdot(2\pi)^{9/2}(a_1a_2a_3)^{-3/2}(a_4a_5a_6)^{-3}\sum_{u,v,w=}^\infty\sum_{\mbox{\scriptsize{even or odd,}}}^\infty\sum_{u,v,w=0}^\infty(3/2+l)(3/2+m)(3/2+n)\\
    &\times(M+1)(M+2)\int_0^\infty J_{3/2}(a_1x)J_{3/2+l}(a_4x)J_{3/2+m}(a_5x)x^{-5/2}\,dx\\
    &\times\int_0^\infty J_{3/2}(a_2x')J_{3/2+m}(a_5x')J_{3/2+n}(a_6x')x^{-5/2}\,dx'\\
    &\times\int_0^\infty J_{3/2}(a_3x'')J_{3/2+l}(a_4x'')J_{3/2+n}(a_6x'')x^{-5/2}\,dx''\\
\end{split}
\end{equation}
\end{widetext}

%figure: B4
\begin{figure}[b]
\centering
\includegraphics[width=8.5cm,trim={0.8cm 0cm 0 0.8cm},clip]{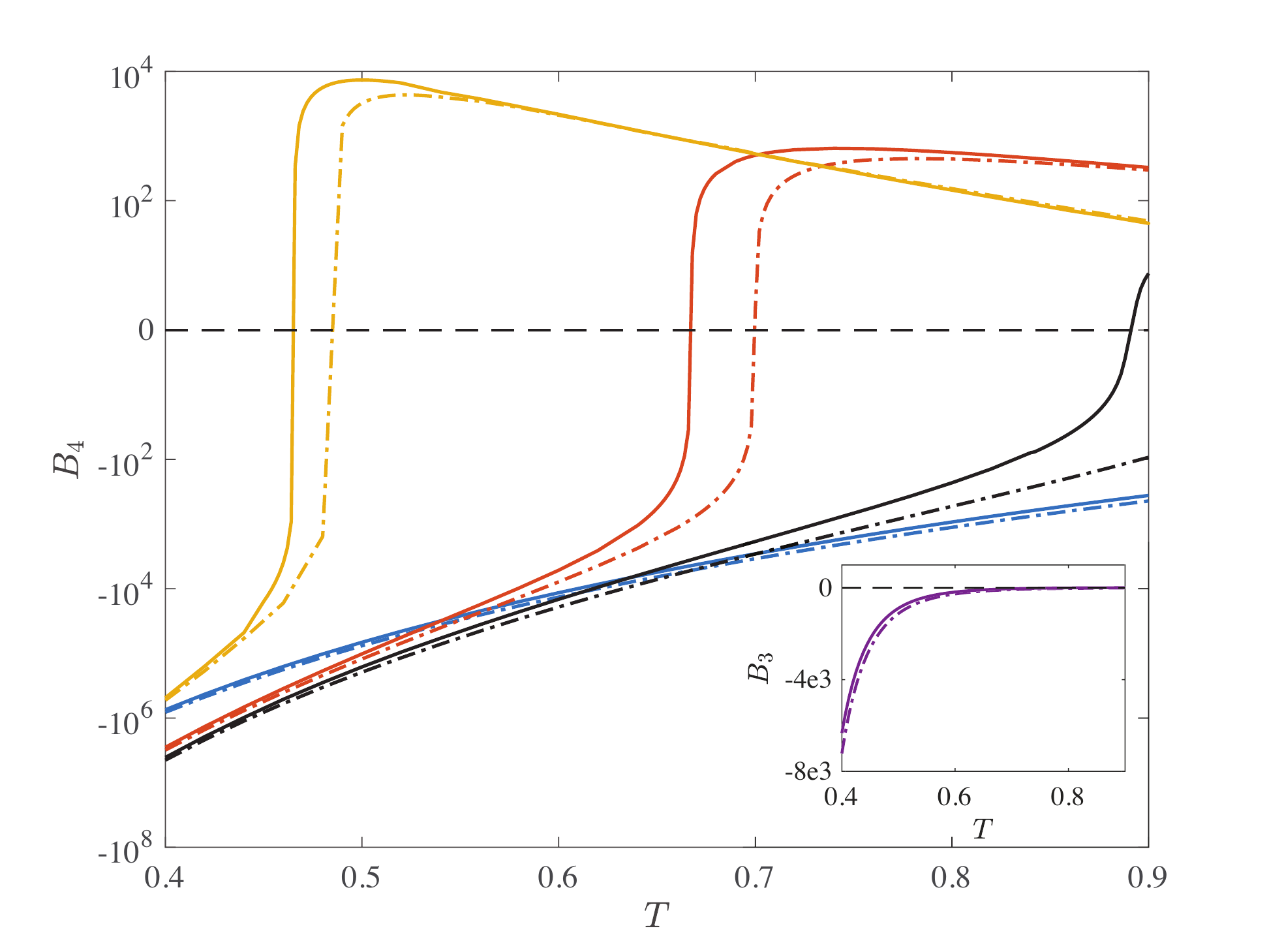}
\caption{Contributions of $B_4(T)$ for Eq.~\eqref{eq:hamiltonian} with $\kappa = 0$ (dashed lines) and $\kappa = 0.025$ (solid lines), from top left to right bottom are $B_4(\XBox)$ (yellow), $B_4(\boxbslash)$ (orange), overall $B_4$ (black) and $B_4(\Square)$ (blue). (Inset) $B_3$ remains negative over the relevant $T$ range.}
\label{fig:b4}
\end{figure}

In practice, standard numerical integrators struggle with  Eq.~\eqref{eq:b4_3_I}, and obtaining analytical solutions of Bessel integrals is beyond our degree of mathematical sophistication. Instead, in $d=1$ we solve Eq.~\eqref{eq:b4_3_I} using Mathematica, and in $d=3$ we numerically evaluate Eq.~\eqref{eq:b4_3} by Monte Carlo integration.

Figure~\ref{fig:b4} compares $B_4$ for our SALR model and for a purely attractive system with $\kappa=0$. It shows that $B_4(\Square)$ is always negative over the temperature range of interest, but that the second and third terms are negative at low $T$ and positive at relatively high $T$. The point at which they change sign shifts to lower $T$ as repulsion increases. Combining all three terms, $B_4$ for the SALR model considered here changes from negative to positive as $T$ increases, around $T=0.89$, while that of purely attraction model remains negative. Interestingly, this temperature roughly corresponds to the emergence of the local minimum in $h(\rho)$ and of the cluster peak in $s\Pi(s)$, where dimers and trimers turn into subclusters (Fig.~\ref{fig:phase}). Hence, even though the equation of state fails to capture clustering directly, it nevertheless hints at the instability of the homogeneous fluid toward it. The consideration of higher-order coefficients might help refine this hypothesis, but falls beyond the scope of the current work.

\bibliographystyle{aip}
\bibliography{main}

\end{document}